\documentclass[a4paper,11pt]{article}
\usepackage{jheppub} 

\preprint{JLAB-THY-23-3818, YITP-SB-2023-08}
\title{Is infrared-collinear safe information all you need \\ for jet classification?}

\usepackage[utf8]{inputenc}
\usepackage[T1]{fontenc}
\usepackage{lmodern}
\usepackage{microtype}
\usepackage{mathtools}

\usepackage{placeins}

\usepackage{amsmath}
\DeclareMathOperator*{\argmax}{arg\, max}

\usepackage{amsmath}
\usepackage{amsfonts}
\usepackage{graphicx}
\usepackage{subcaption}
\usepackage{mwe}
\usepackage{multirow}
\hypersetup{
  colorlinks=true,
  citecolor=blue,
  linkcolor=blue,
  urlcolor=blue
}

\usepackage{color}
\definecolor{darkblue}{rgb}{0,0,0.5}

\author[1,2]{Dimitrios Athanasakos,}
\affiliation[1]{C.N. Yang Institute for Theoretical Physics, Stony Brook University, Stony Brook, NY 11794, USA}
\affiliation[2]{Department of Physics and Astronomy, Stony Brook University, Stony Brook, NY 11794, USA}
\emailAdd{dimitrios.athanasakos@stonybrook.edu}

\author[3]{Andrew J.~Larkoski,}
\affiliation[3]{Department of Physics and Astronomy, University of California, Los Angeles, CA 90095, USA}
\emailAdd{larkoski@ucla.edu}

\author[4,5]{James Mulligan,}
\affiliation[4]{Nuclear Science Division, Lawrence Berkeley National Laboratory, Berkeley, California 94720, USA}
\affiliation[5]{Physics Department, University of California, Berkeley, CA 94720, USA}
\emailAdd{james.mulligan@berkeley.edu}

\author[4]{Mateusz P\l osko\'n,}
\emailAdd{mploskon@lbl.gov}

\author[1,2,6,7]{Felix Ringer}
\affiliation[6]{Department of Physics, Old Dominion University, Norfolk, VA 23529, USA}
\affiliation[7]{Thomas Jefferson National Accelerator Facility, Newport News, VA 23606, USA}
\emailAdd{fmringer@jlab.org}

\abstract{
Machine learning-based jet classifiers are able to achieve impressive tagging performance in a variety of applications in high-energy and nuclear physics.
However, it remains unclear in many cases which aspects of jets give rise to this discriminating power, and whether jet observables that are tractable in perturbative QCD such as those obeying infrared-collinear (IRC) safety serve as sufficient inputs.
In this article, we introduce a new classifier, 
Jet Flow Networks (JFNs), in an effort to address the question of whether IRC unsafe information provides additional discriminating power in jet classification. 
JFNs are permutation-invariant neural networks (deep sets) that take as input the kinematic information of reconstructed subjets.
The subjet radius and a cut on the subjet's transverse momenta serve as tunable hyperparameters enabling a controllable sensitivity to soft emissions and nonperturbative effects. We demonstrate the performance of JFNs for quark vs. gluon and Z vs. QCD jet tagging.
For small subjet radii and transverse momentum cuts, the performance of JFNs is equivalent to the IRC-unsafe Particle Flow Networks (PFNs), demonstrating that infrared-collinear unsafe information is not necessary to achieve strong discrimination for both cases.
As the subjet radius is increased, the performance of the JFNs remains essentially unchanged until physical thresholds that we identify are crossed.
For relatively large subjet radii, we show that the JFNs may offer an increased model independence with a modest tradeoff in performance compared to classifiers that use the full particle information of the jet. These results shed new light on how machines learn patterns in high-energy physics data.}

\begin{document}

\maketitle

\section{Introduction}

Jets are highly energetic and collimated groups of particles observed in the detectors of high-energy scattering experiments such as the Large Hadron Collider (LHC)~\cite{Larkoski:2017jix,Asquith:2018igt,Marzani:2019hun}.
Jets arise from the fragmentation of highly energetic quarks and gluons, 
which themselves can arise from the decay of unstable particles such as the Higgs boson. 
Classifying the origins of jets, such as 
quark vs. gluon initiated jets, QCD vs. boosted $Z/W$ jets, and QCD vs. boosted top jets~\cite{Lonnblad:1990bi,deOliveira:2015xxd,
Komiske:2016rsd, Komiske:2017aww, Kasieczka:2017nvn,Louppe:2017ipp,Datta:2017rhs,Komiske:2018cqr,Qu:2019gqs,Kasieczka:2020nyd,Dreyer:2020brq,Top_taggers_review, Butter:2017cot, Chen:2019uar,Araz:2021wqm, Gong:2022lye, Schwartz:2021ftp,Khot:2022aky, Lin:2018cin, Khosa:2021cyk, Nakai:2020kuu}  
is crucial to disentangle the various processes occurring at collider experiments and perform searches for physics beyond the Standard Model.

Jet classification algorithms have been developed based on multivariate combinations of jet substructure observables as well as using machine learning methods. Machine learning based jet classifiers significantly outperform traditional multivariate jet taggers that utilize a limited number of observables, since they are able to leverage the full information in the jet. However, machine learning based classifiers often have the drawback that they are not calculable by analytical methods. 
Efforts to address this have been an active area of research, such as enforcing Infrared-Collinear (IRC) safety~\cite{CTEQ:1993hwr} in the network architecture~\cite{EnergyFlowPolynomials:2017aww, Romero:2021qlf, Konar:2021zdg, Larkoski:2019nwj, Choi:2018dag}
or by finding optimal ways to reduce the amount of information provided as input to neural networks ~\cite{Lai:2021ckt, Feature_selection_Shih, Thaler_human_readable_space, Bogatskiy:2020tje, Lu:2022cxg}. IRC safety is a frequently used guiding principle in high-energy physics for constructing suitable observables. It is often stated as: ``An observable is IRC safe if it is insensitive to infinitesimally soft or exactly collinear emissions''~\cite{Ellis:1996mzs}. It is a defining feature of observables that allows for calculations in QCD where divergences cancel at each order in perturbation theory~\cite{Ellis:1993tq,Larkoski:2017jix}. We note one has to distinguish between the IRC safety of an observable, i.e. the calculability or traceability of an observable within perturbative QCD, and the question about the relevance of nonperturbative effects. While both questions are important, we will primarily focus on IRC safety throughout this work.

In order to increase the interpretability of machine learning based classifiers, a complete IRC-safe basis of jet substructure observables was introduced in Refs.~\cite{Datta:2017rhs,Datta:2017lxt,Datta:2019ndh} based on $N$-subjettiness observables~\cite{Thaler:2010tr,Thaler:2011gf}. These observables capture the momentum and relative angles of emissions inside the jet. The set of $N$-subjettiness observables is then used as input to a machine learning algorithm for jet classification. While the complete basis of IRC-safe observables is large ($3M-4$ for $M$ particles in the jet), it was found that the performance of classifiers saturates quickly with a relatively small number of observables. Another set of observables, Energy Flow Polynomials (EFPs), was developed as a linear and IRC-safe basis of jet substructure observables in Ref.~\cite{EnergyFlowPolynomials:2017aww}. 

Interestingly, it was found that while the performance of classifiers based on complete sets of observables saturates, in most cases there remains a performance gap between classifiers with IRC-safe inputs (Sudakov safe classifiers) and IRC-unsafe classifiers that make use of the full information content of the particles inside the jet. Examples of such IRC-unsafe classifiers include architectures based on deep sets~\cite{Komiske:2018cqr}, point clouds~\cite{Qu:2019gqs, Mikuni:ABCNet} and transformers~\cite{Qu:2022mxj}. This performance gap has been observed for a variety of jet classification tasks, including QCD vs. $W/Z$ and $H$ jets~\cite{Lu:2022cxg}, and $pp$ vs. $AA$ jets~\cite{Lai:2021ckt}. For quark vs. gluon tagging, it was found that the IRC-safe EFPs can match the performance of PFNs when only momentum information of particles in the jet is considered~\cite{Komiske:2018cqr}. Several efforts have been made to quantify the gap, with the aim to gain new insights into fundamental QCD dynamics. There are several possible explanations for the observed performance gap:
\begin{itemize}
    \item IRC-unsafe classifiers may be able to make use of the very soft information content of jets, which is difficult to access with IRC-safe observables.
    \item IRC-unsafe classifiers such as PFNs take as input the exact position information of the particles inside the jet, whereas IRC-safe observables can only capture the information of relative distances. It is possible that existing machine learning algorithms can make more efficient use of position information.
    \item The specific form of the IRC-safe observables may not be optimal for classification tasks and there may be other sets of observables that could perform better.
\end{itemize}
With this question in mind, we introduce in this work a new machine learning-based jet classifier, Jet Flow Networks (JFNs)\footnote{In analogy to Particle Flow Networks (PFNs)~\cite{Komiske:2018cqr}.}, which take as input the energy and position of reclustered subjets instead of individual particles. 
JFNs allow for soft and collinear emissions to be clustered into subjets making the input IRC-safe and the resulting classifier generally Sudakov safe~\cite{Larkoski:2015lea, Larkoski:2013paa}. Sudakov safe observables are not IRC-safe, and as such are not defined at any order in the strong coupling $\alpha_s$ and yet have finite cross sections when all-orders effects are included. This happens because the Sudakov form factor regulates the infrared divergences. An example of such an observable is the ratio formed from two different angularities~\cite{Berger:2003iw} measured on the same jet. However, different than the $N$-subjettiness or the EFP basis of observables, position information is used instead of having (indirectly) access only to relative distances between emissions (or subjets) inside the jet. We note that we do not consider quark flavor tagging in this work which requires IRC-unsafe information, see e.g. Refs.~\cite{Fraser:2018ieu,Lee:2022kdn,Kang:2023ptt}. JFNs are closely related to Particle Flow Networks (PFNs)~\cite{Komiske:2018cqr} and Energy Flow Networks (EFNs)~\cite{Komiske:2018cqr}, which will be elaborated on in section~\ref{sec:deep_sets}. In the limit of a vanishing subjet radius, where every subjet contains only a single hadron, and in the limit of a vanishingly small cut on the transverse momenta of the subjets, JFNs are identical to PFNs.
The radius of the reclustered subjets in JFNs, as well as the cut on the transverse momenta, can be used to dial in nonperturbative information allowing for a smooth transition to IRC-unsafe classifiers.  
As such, JFNs complement the existing family of permutation-invariant networks in particle physics.
As for PFNs and EFNs, we will utilize machine learning algorithms for JFNs based on a permutation invariant deep set architecture~\cite{DBLP:journals/corr/ZaheerKRPSS17,DBLP:journals/corr/abs-1901-09006,JMLR:v21:19-322, Komiske:2018cqr}.

In this paper, we will explicitly study classification tasks of quark vs.~gluon jets and jets from QCD vs.~jets from boosted hadronic decays of $Z$ bosons.  The particular IRC (un)safety of the likelihood ratios for these tasks has been studied previously.  It is expected that the likelihood ratio for quark vs.~gluon jet discrimination is IRC safe~\cite{Larkoski:2019nwj}, which has been validated in previous machine learning studies~\cite{Romero:2021qlf,Konar:2021zdg,Dreyer:2021hhr}.  The argument for IRC safety is that the $N$-body phase space can be spanned by additive, IRC safe observables~\cite{Banfi:2004yd} and so probability distributions for quarks and gluons will in general take the form of a Sudakov factor near the infrared regions of phase space.  The rate of suppression of emissions is controlled by the corresponding color Casimirs, and, because gluons carry more color than quarks, the likelihood ratio itself vanishes at all infrared boundaries.  Therefore, all fixed-order divergences are mapped to the same value of the likelihood ratio, namely 0, and so the classifier is IRC safe. This explains why EFPs, which are an IRC-safe classifier, can match the performance of the IRC-unsafe PFNs. By contrast, the likelihood ratio for QCD jets vs.~$Z$ jets is expected to only be Sudakov safe, as optimal observables for general one- vs.~two-prong discrimination take the form of the ratio of IRC safe observables~\cite{Almeida:2008tp,Thaler:2010tr,Larkoski:2013eya,Larkoski:2014gra}.  Ratios of IRC safe observables are in general themselves not IRC safe \cite{Soyez:2012hv}.

To separately study soft and collinear safety, we consider both a finite subjet radius (collinear safety) and a low momentum cut $p_T^{\rm soft}$ on the transverse momentum of the subjets (soft safety). The main result of our work will be to show that the JFNs based on IRC-safe input achieve the same classification performance as PFNs. The JFNs represent the first example of a classifier based on IRC-safe inputs that achieve equivalent performance on several classification tasks as its IRC-unsafe counterpart. In addition, the machine learning architecture is equivalent to PFNs, which allows for one-to-one comparisons. The exact value of the subjet radius where the PFN performance is matched depends on the classification task at hand. Therefore, different than the classifiers based on complete IRC-safe sets of observables, JFNs constitute a ``gapless'' classifier indicating that the very soft aspects of jets are in fact not relevant for typical classification tasks at collider experiments. This answers in part the question about the features that are relevant for the performance of classifiers in high-energy physics. Throughout this work, PFNs are taken as a reference, but other permutation invariant classifiers such as GNNs, transformers (equivalent to a fully connected graph), and point clouds could equally well be trained on particles or reclustered subjets. 

\begin{figure*}[!t]
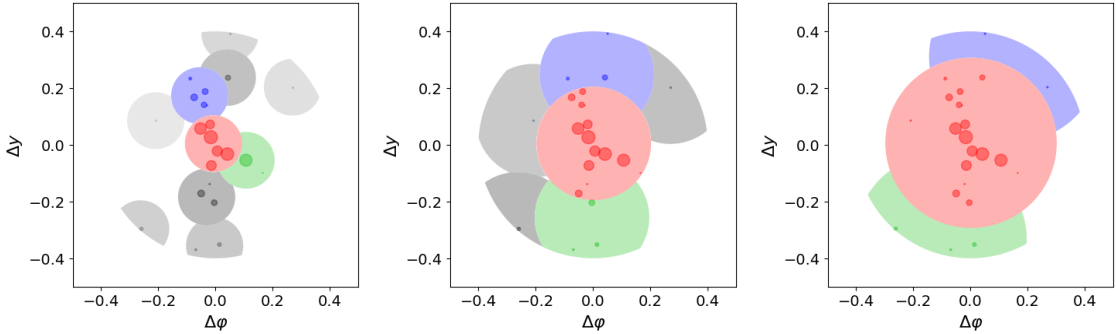

    \includegraphics[scale=0.40]{r01.pdf}
    \includegraphics[scale=0.40]{r02.pdf}
    \includegraphics[scale=0.40]{r03.pdf}
\caption{Illustration of a QCD jet with $p_T=100$~GeV and radius parameter $R=0.4$ reclustered into subjets for subjet radii $r=0.1$ (left), $r=0.2$ (middle), and $r=0.3$ (right). We use the inclusive anti-$k_T$ algorithm to identify the initial jet and the subjets. 
Particles are represented by small filled circles with radii proportional to the particle's transverse momentum in the $\Delta y$ vs. $\Delta\varphi$ plane, where $\Delta\varphi=\varphi^{\mathrm{particle|subjet}}-\varphi^{\mathrm{jet}}$ is the azimuthal angle with respect to the jet axis and $\Delta y = \Delta y^{\mathrm{particle|subjet}} - \Delta y^{\mathrm{jet}}$ is rapidity distance to the jet axis. 
Subjets are shown with larger colored areas where red marks the leading subjet, green marks the second leading subjet, blue marks the third leading jet, and shades of gray represent subjets with lower longitudinal momentum fraction $z=p_T^{\rm subjet}/p_T$ with intensity proportional to $z$.~\label{fig:subjets_illustration}}
\end{figure*}

In addition to shedding light on the role of IRC-safe information, JFNs allow for new insights into the physics of jet tagging and may lead to various future applications at high-energy collider experiments. By studying the performance of the JFNs as a function of the subjet radius and the jet transverse momentum, we are able to identify the relevant physical scales of different classification tasks. For example, for $Z$ vs. QCD-jet tagging, we find that the subjet scale $p_T r$ is sensitive the opening angle between the boosted hadronic decay products of the $Z$-boson. Second, we explore the generalization capability of JFNs to unseen data, which is crucial when deploying a classifier trained on simulations to experimental data. Due to the clustering of collinear and soft emissions into subjets, the resulting JFNs are relatively insensitive to the detailed modeling of the infrared (IR) physics that is often poorly understood. This raises the possibility to use JFNs to trade performance for generalizability by adjusting the number of reconstructed subjets. Lastly, we expect that subjets can be measured well in heavy-ion collisions despite the large fluctuating background. See Ref.~\cite{ALICE:2022vsz} for recent measurements of the energy spectrum of inclusive and leading subjets by the ALICE Collaboration. See also Ref.~\cite{Chen:2021uws}.

The remainder of this paper is organized as follows. In section~\ref{sec:subjets}, we introduce the subjet basis and discuss differences between inclusive and exclusive subjet reconstruction algorithms. In section~\ref{sec:deep_sets}, we introduce the permutation invariant machine learning algorithms that take the kinematic information of the reconstructed subjets as input and in section~\ref{sec:datasets}, we briefly discuss the data sets used for different classification tasks used in this work. In section~\ref{sec:applications}, we present numerical results for the classification performance of JFNs for quark vs. gluon and $Z$ vs. QCD jets. In particular, we show that JFNs match the PFN performance for a finite subjet radius and a finite cut on the transverse momenta of the subjets in both test cases. Based on these results, we describe in section~\ref{sec:learning} that the machine learning algorithm is sensitive to different physical scales, which it can effectively learn. In section~\ref{sec:generalization}, we investigate the tradeoff between performance and generalizability of the JFNs. In section~\ref{sec:conclusions}, we draw conclusions and present an outlook.

\section{The subjet basis~\label{sec:subjets}}

In this section, we describe the reconstruction of subjets that will serve as the input to the machine learning classifier. The initial jet is identified using the anti-$k_T$ algorithm~\cite{Cacciari:2008gp} and jet radius parameter $R$. In order to utilize the substructure of jets, we then recluster the jet constituents into subjets. We consider two approaches for the subjet reconstruction: inclusive anti-$k_T$ subjets  and the exclusive $k_T$ subjets~\cite{Catani:1993hr,Cacciari:2011ma}. Note that we do not consider exclusive anti-$k_T$ jets as this clustering procedure does not lead to a ``physical'' clustering tree. In both cases, soft and collinear emissions are first clustered into subjets. For a finite subjet radius and an additional cut on the subjet's transverse momentum $p_T^{\rm soft}$, the input to the classifier is IRC safe. In this sense, subjets serve as a useful tool for throttling or controlling the input data to the machine in a way that is theoretically interpretable in perturbative QCD.

First, we consider inclusive subjets reconstructed with the  anti-$k_T$ algorithm using the $E$-recombination scheme~\cite{Blazey:2000qt} and a fixed subjet radius $r<R$. This approach fixes the maximally allowed size of the reconstructed subjets but the number of subjets varies for each jet. We illustrate the distribution of subjets in the $\eta$-$\phi$ plane for three different subjet radii in Fig.~\ref{fig:subjets_illustration}. As $r$ is increased, the central subjet contains a large fraction of particles. Second, we consider subjets reconstructed with the exclusive $k_T$ algorithm and the $E$-recombination scheme. Particles are clustered with the $k_T$ algorithm until a fixed number of subjets $N$ is obtained. Different than in the case of inclusive subjets, the number of identified subjets is fixed but their size varies jet-by-jet. The $N$ subjets span the full information content of the $N$ most resolved emissions inside the jet analogous to the $N$-subjettiness basis developed in Refs.~\cite{Datta:2017rhs,Datta:2017lxt,Datta:2019ndh}. An alternative approach to the exclusive $k_T$ algorithm is to identify subjets with the XCone algorithm~\cite{Stewart:2015waa}. We leave the exploration of this algorithm for future work. By taking the small-$r$ (inclusive subjets) or large-$N$ limit (exclusive subjets), as well as $p_T^{\rm soft}\to 0$, we can study the transition to the nonperturbative regime where eventually, every subjet only contains a single hadron. 

\begin{figure*}[t]
\includegraphics[width=0.5\textwidth]{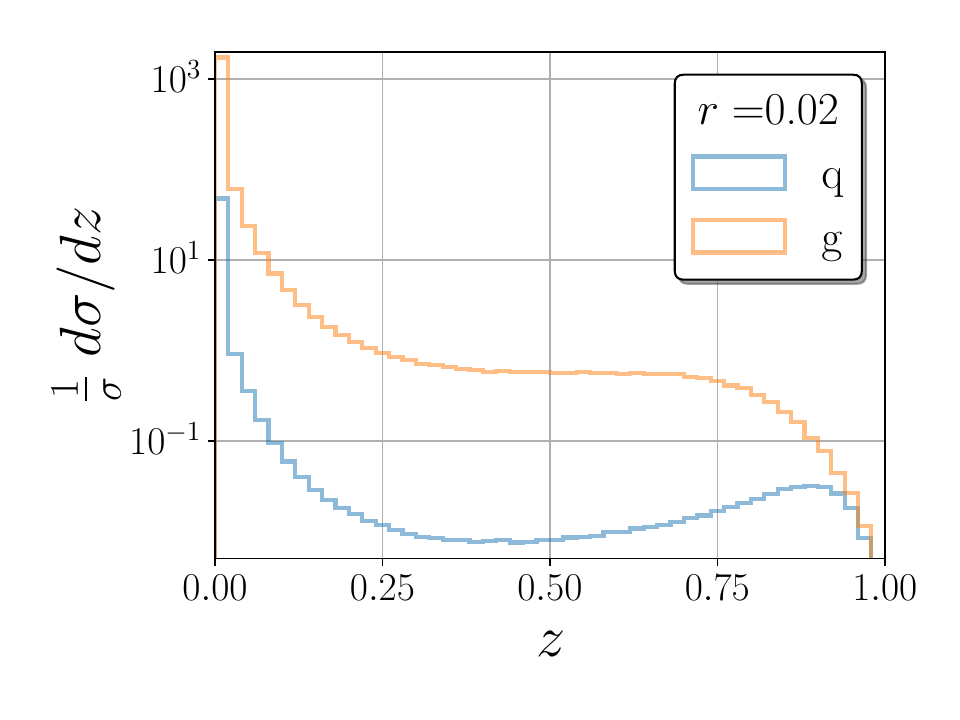}
\hspace*{0.2cm}
\includegraphics[width=0.5\textwidth]{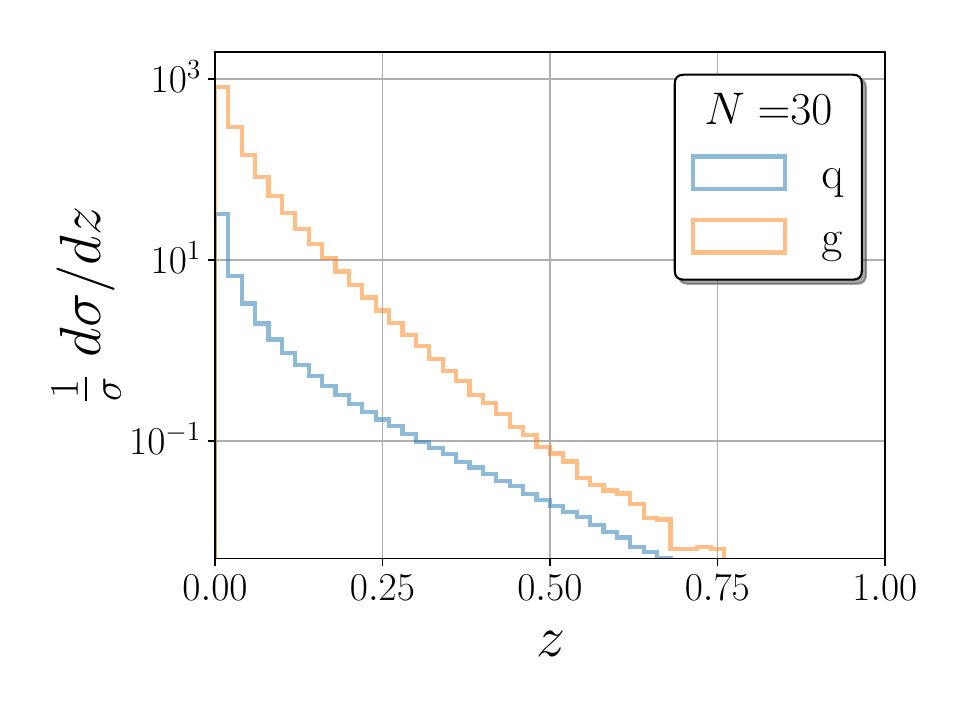}
\caption{The longitudinal momentum distribution of inclusive subjets $z=p_T^{\rm subjet}/p_T$ originating from either a quark (blue) or a gluon (orange) jet simulated with \textsc{Pythia}~\cite{Sjostrand:2014zea}. The jets have $p_T = [500, 550]$~GeV with an average particle multiplicity of 43. We show the distributions for inclusive anti-$k_T$ subjet clustering with $r=0.02$ (left) corresponding to an average of approximately 30 subjets, and for exclusive $k_T$ clustering with a fixed number of $N=30$ subjets (right). See also section~\ref{sec:datasets} for more details. ~\label{fig:subjets_z}}
\end{figure*}

To illustrate the qualitative differences between the two reconstruction methods discussed above, we show as an example the longitudinal momentum distributions of subjets $z=p_T^{\rm subjet}/p_T$ in Fig.~\ref{fig:subjets_z} separately for quark and gluon jets. The details on the data generation are presented in section~\ref{sec:datasets}. Here $p_T$ denotes the initial jet transverse momentum and $p_{T}^{\rm subjet}$ the longitudinal subjet momentum using either the inclusive or exclusive reconstruction method. As an example, we choose $N=30$ for the exclusive reconstruction of subjets and $r=0.02$ for inclusive subjets, which yields a comparable average number of 30 subjets. We observe that the two methods lead to qualitatively different spectra. The inclusive subjet spectrum exhibits a peak (quarks) or plateau (gluons) for intermediate to large values of $z$. In contrast, the spectrum for exclusive subjets only peaks at small values of $z$ and falls off steeply for $z\to 1$. This is due to the fact that for exclusive clustering the $k_T$ algorithm is used, where soft hadrons are clustered first. Only at the end hard emissions are combined, making it unlikely to find a subjet with $z\to 1$ for a fixed value of $N$. We note that for inclusive subjets, the $z$-distributions are qualitatively the same for both the anti-$k_T$ and $k_T$ algorithms. The longitudinal momentum spectrum for inclusive subjets was calculated within perturbative QCD up to next-to-leading logarithmic (NLL) accuracy. See Refs.~\cite{Dasgupta:2014yra,Dai:2016hzf,Kang:2017mda,Neill:2021std}. This close connection to first-principles calculations may allow for an increased understanding of machine learning algorithms in QCD.

From the identified subjets, the kinematic information $(z_i,\eta_i,\phi_i)$ of the each subjet is used as input to the classifiers discussed below. In the limit that $r\rightarrow 0$ (inclusive subjets) or $N\rightarrow \infty$ (exclusive subjets), the subjet basis becomes equivalent to the set of particle four-vectors of the jet, and the classifier can make use of the full information content of the jet. 
The subjet basis therefore provides a means to limit the information supplied to the classifier, by using $r>0$ or $N<\infty$.

\section{Jet Flow Networks (JFNs): Deep sets of subjets~\label{sec:deep_sets}}

In this section, we describe the permutation invariant neural networks that use the kinematic information of subjets as input to perform binary classification tasks. As introduced above, we refer to the machine learning architecture and the pre-processing step of clustering particles into subjets as JFNs.

The reconstructed subjets discussed in the previous section do not have an inherent ordering. Therefore, permutation-invariant neural networks are a natural choice to perform classification tasks that take as input the kinematic information of subjets. In Refs.~\cite{DBLP:journals/corr/ZaheerKRPSS17,DBLP:journals/corr/abs-1901-09006,JMLR:v21:19-322} deep sets were introduced as a permutation invariant neural network. In the context of particle physics deep sets were first discussed in Ref.~\cite{Komiske:2018cqr} as Particle Flow Networks (PFNs) that take as input the information of individual particles. A permutation invariant classifier $f$, which takes as input the subjet four-momenta $p_i$ satisfies $f(p_1,\ldots,p_N)=f(p_{\pi(1)},\ldots,p_{\pi(N)})$. Here $\pi$ denotes the permutation operator. Following Ref.~\cite{DBLP:journals/corr/ZaheerKRPSS17}, we can write the classifier $f$ as
\begin{equation} \label{eq:PFN}
    f(p_1,\ldots,p_n)=F\bigg(\sum_{i=1}^N \Phi_i(p_i)\bigg) \,, 
\end{equation}
where $F,\ \Phi$ are neural networks and, as an intermediate step, we sum over all reconstructed subjets $N$. The first neural network $\Phi: \mathbb{R}^4\to \mathbb{R}^l$ takes as input the individual subjet four momenta and maps it to an $l$-dimensional latent space. For massless subjets, we can write the individual four vectors in terms of $(z_i,\eta_i,\phi_i)$. Here $z_i$ is the subjet's longitudinal momentum fraction, see Fig.~\ref{fig:subjets_z}, and $(\eta_i,\phi_i)$ denote its coordinates in the rapidity-azimuth plane. We note that further information can be included in the per-subjet mapping such as the jet mass or the jet charge~\cite{Waalewijn:2012sv,Krohn:2012fg}, analogous to e.g. particle identification (PID) for PFNs.
We leave quantitative studies of the impact of these additional features for future work. The summation in Eq.~\eqref{eq:PFN} ensures that the classifier $f$ is invariant under permutations of the input variables. The second neural network $F: \mathbb{R}^l\to \mathbb{R}$ is a map from the latent space where the summation operation is performed to the final classification score. Note that the classifier architecture in Eq.~\eqref{eq:PFN} can accommodate both a fixed number $N$ of subjets (exclusive subjets) and input with variable length (inclusive subjets).
\begin{center}
\begin{table*}[!t]
\centering
\begin{tabular}{|| c | c | c | c ||} 
 \hline
   & PFN~\cite{Komiske:2018cqr} & JFN & EFN~\cite{Komiske:2018cqr} \\ [0.5ex] 
 \hline\hline
 Input & particle 3-momenta & subjet 3-momenta & particle 3-momenta \\ 
 \hline
 Classifier & IRC unsafe & Sudakov safe & IRC safe \\
 \hline
\end{tabular}
\caption{Overview of different classifiers based on permutation invariant neural networks.~\label{tab:classifiers}}
\end{table*}
\end{center}

We refer to the deep set classifier based on subjets in Eq.~\eqref{eq:PFN} as JFNs. 
The JFNs are a family of classifiers due to the dependence on the continuous parameter $r$ in the case of inclusive clustering or on $N$ in the case of exclusive clustering, in which case the clustering is performed until $N$ subjets remain.
Since the JFN takes subjet information as input, the resulting classifier is generally Sudakov safe~\cite{Larkoski:2013paa}. We summarize the different aspects of permutation invariant network architectures based on deep sets in table~\ref{tab:classifiers}.
Since the JFNs are Sudakov safe, they constitute an intermediate point between IRC-unsafe PFNs and IRC-safe EFNs. In the limit of $r\to 0$ (inclusive subjets) or the large-$N$ limit (exclusive subjets) and $p_T^{\rm soft}\to 0$, we recover the PFN classifier. 

We note that one can think of the input to the classifier as a multi-differential cross section or multi-variate probability distribution. To avoid for example regions of negative cross sections, the corresponding perturbative calculation requires the all-order resummation of sufficiently many logarithmic corrections. The resummed calculation then needs to be matched to fixed-order calculations. The classifier performs a highly nontrivial marginalization over the multi-variate probability distribution. Since all order resummations are required, the final result of the classifier is Sudakov but not necessarily IRC safe.

\section{Data sets~\label{sec:datasets}}

In this work, we will consider JFNs for two exemplary binary classification tasks in high-energy physics. First, we consider quark vs. gluon jet classification and, second, $Z$ vs. QCD jet classification. For the quark vs. gluon case, we make use of the data set in Ref.~\cite{Zenodo:EnergyFlow:Pythia8QGs}, which consists of 2M jets with transverse momentum $p_T = [500, 550]$~GeV, rapidity $|\eta|<1.7$, jet radius parameter $R=0.4$, and center-of-mass energy $\sqrt{s}=14$~TeV. We will make use of both the data set generated with \textsc{Pythia}~\cite{Sjostrand:2014zea} and \textsc{Herwig}~\cite{Bellm:2019zci}. In order to explore the dependence on the jet transverse momentum, we also generate two additional data sets consisting of 500k jets each with transverse momentum $p_T = [300, 350]$~GeV and $[1000,1050]$~GeV, respectively. The underlying processes are:  
$q\bar{q} \rightarrow Z(\rightarrow \nu \bar{\nu}) + g$ and $q\bar{q} \rightarrow Z(\rightarrow \nu \bar{\nu}) + (uds)$ analogous to Ref.~\cite{Zenodo:EnergyFlow:Pythia8QGs}.
For the $Z$ vs. QCD-jet case, we generate 500k jets for three different bins of jet transverse momentum, [300, 350] GeV, [500, 550] GeV and [1000, 1100] GeV with a jet mass $m_j=[45,135]$ GeV. The radius parameter is $R=0.8$, the rapidity cut is  $|\eta|<1.7$ and the samples are generated using \textsc{Pythia} at $\sqrt{s}=14$~TeV. Jets arising from $Z$ bosons are identified by requiring that the leading $Z$ boson is in the catchment area of the jet as extracted from the kinematics of the events at the particle level before hadronization with a $Z$-jet distance from the jet axis less than $R/2$. A similar tagging procedure is performed to differentiate between quark and gluon jets in the QCD sample. The tag is based on the leading parton within the catchment area of the jet. However, to strengthen the parton-jet association, we use parton-level kinematics injected into the hadron-level event using so-called ghost particles ($p_{T}=10^{-5}$~GeV) that do not affect the jet reconstruction but allow for efficient tagging after the jet finding step.

\begin{figure}[t!]
    \centering
    \includegraphics[width=0.495\textwidth]{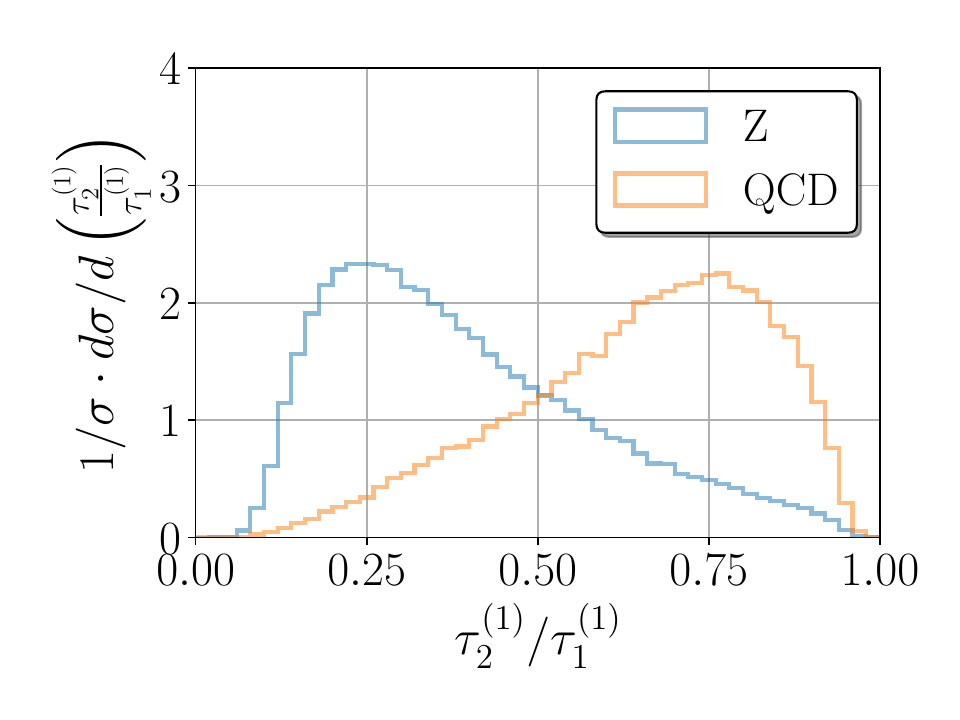} \includegraphics[width=0.495\textwidth]{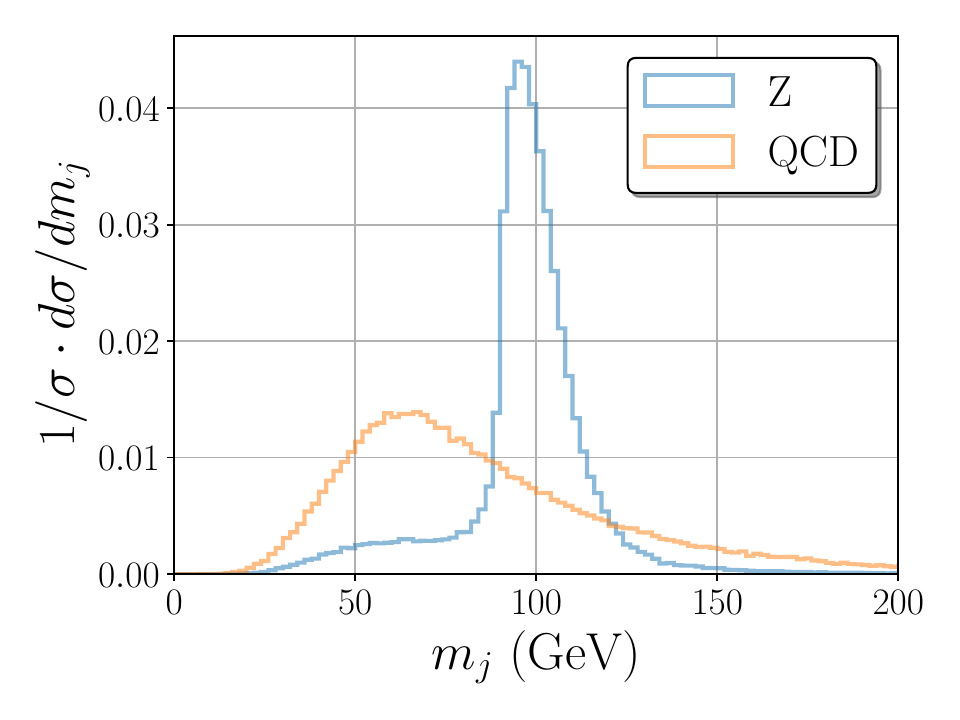} 
    \caption{$N$-subjettiness ratio $\tau_2^{(1)} /\tau_1^{(1)}$ (left) and the jet mass distribution (right) for QCD and $Z$ jets with $p_T = [500, 550]$~GeV. The N-subjettiness axes were identified using the one pass $k_T$ clustering algorithm.~\label{fig:tau2tau1}}
\end{figure}

The substructure of QCD jets is generally single-pronged, whereas the decay products of a $Z$ cause the corresponding jets to have two prongs. The ratio of $N$-subjettiness observables~\cite{Thaler:2010tr,Thaler:2011gf,Stewart:2010tn,Napoletano:2018ohv} is sensitive to the number of prongs inside a jet. In order to define the $N$-subjettiness, a given number of $N$ axes are identified inside the jet using the exclusive $k_T$ algorithm. The $N$-subjettiness variables $\tau_N^{(\beta)}$ measure the radiation along these axes and are defined as
\begin{equation}
    \tau_{N}^{(\beta)}=\frac{1}{p_T} \sum_{i \in \mathrm{jet}} p_{T i} \min \left\{R_{1 i}^{\beta}, R_{2 i}^{\beta}, \ldots, R_{N i}^{\beta}\right\} \,.
\end{equation}
Here the $p_{Ti}$ of each particle $i$ is weighted by its distance $R_{ji}$ to the closest axis $j$ raised to the power $\beta>0$, which is a tunable parameter. For jets that are more like a single-prong jet, the variable $\tau_2$ will peak at smaller values compared to $\tau_1$, whereas for two-prong like jets the variable $\tau_2$ takes similar values compared to $\tau_1$ (by construction, $\tau_{n+1}\leq \tau_n$). To illustrate the qualitative differences between QCD jets and $Z$ jets, we plot in Fig.~\ref{fig:tau2tau1}, the result for the ratio $\tau_2^{(1)}/\tau_1^{(1)}$, which shows the expected separation of the two jet samples (left panel), and the jet mass $m_j$ distribution for QCD and $Z$ jets (right panel). For all classification tasks, the training/validation/test split is 80\%/10\%/10\%.

\section{JFN performance: gapless jet classification~\label{sec:applications}}

In this section, we will present results for the performance of the JFN classifier. In the two subsections, we separately study the JFN dependence on the subjet radius $r$ and the transverse momentum cutoff $p_T^{\rm soft}$, which are associated with collinear and soft safety, respectively. We will observe that the maximum performance is obtained for finite values $r$ and $p_T^{\rm soft}$.

\subsection{Collinear safety}

We start by studying JFNs for different values of the subjet radius $r$. In the limit $r\to 0$, every subjet contains only a single hadron, and the PFN performance is recovered. Throughout this section, we consider $p_T^{\rm soft}=0$~GeV. In the next subsection, we consider the JFN performance for finite values of the transverse momentum cutoff. We consider two exemplary binary classification tasks in high-energy physics: quark vs. gluon and $Z$ vs. QCD jet identification.

In order to implement the permutation-invariant neural networks, we parametrize the functions $\Phi$ and $F$ in Eq.~(\ref{eq:PFN}) in terms of DNNs, using the \texttt{EnergyFlow} package~\cite{Komiske:2018cqr} with \texttt{Keras}~\cite{chollet2015keras}/\texttt{TensorFlow}~\cite{tensorflow2015-whitepaper}. For $\Phi$ we use two hidden layers with 100 nodes each and a latent space dimension of $d=256$. For $F$ we include three layers with 100 nodes each. For each dense layer, we use the ReLU activation function~\cite{nair2010rectified} and we use the softmax activation function for the final output layer of the classifier. We train the neural networks using the Adam optimizer~\cite{Kingma2015AdamAM} with learning rates ranging from $10^{-3}$ to $10^{-4}$. We use the binary cross entropy loss function~\cite{https://doi.org/10.1111/j.2517-6161.1958.tb00292.x} and train for 60 epochs with a batch size of 256 and a patience parameter of 8 for early stopping. We find no significant changes in performance when changing the size or number of the layers, latent space dimension, learning rate, and batch size by factors of 2-5.
Following Ref.~\cite{Komiske:2018cqr}, we perform a preprocessing step to simplify the training process: we use the rescaled momentum fractions $z_i$ and center the rapidity and azimuthal angles $\eta_i,\phi_i$ of the particles in the jet with respect to the jet direction.

We quantify the performance of the different classifiers in terms of the Receiver Operator Characteristic (ROC) curve and the Area Under the Curve (AUC) of ROC curve. The ROC curve is the cumulative distribution function of the true positive rate vs. the false positive rate of a classifier as the decision threshold is varied. The AUC takes values between 0.5 and 1, where 0.5 (1) corresponds to a random (perfect) binary classifier. We estimate the statistical uncertainty of the AUC by training the deep sets four times for each choice of the subjet radius $r$ and using the standard deviation as the uncertainty.

\begin{figure}
    \centering
    \includegraphics[width=0.485\textwidth]{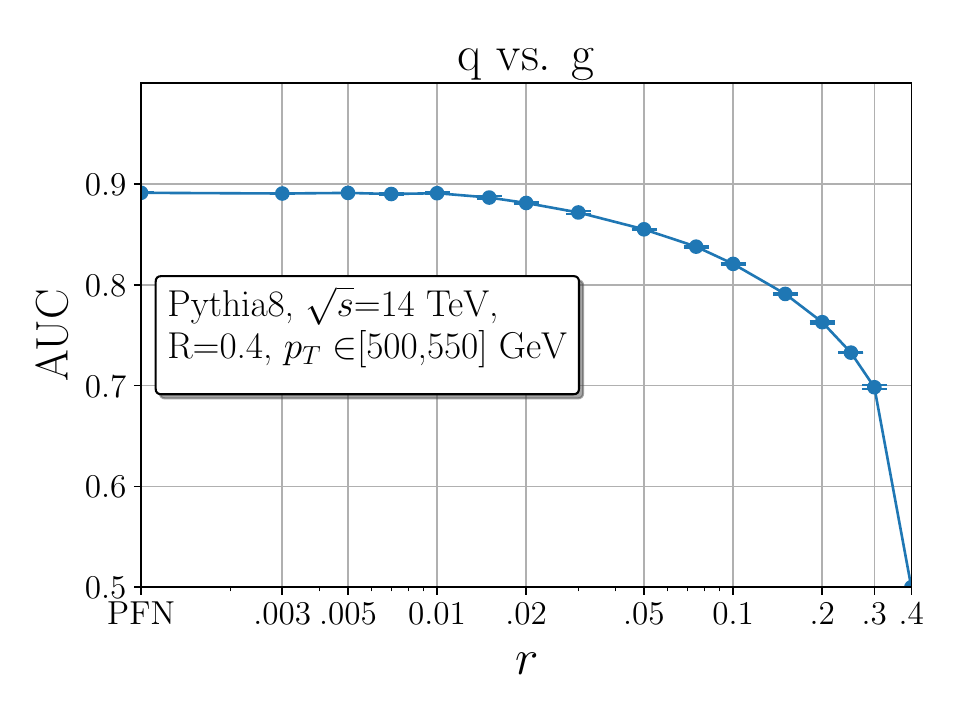} 
    \includegraphics[width=0.485\textwidth]{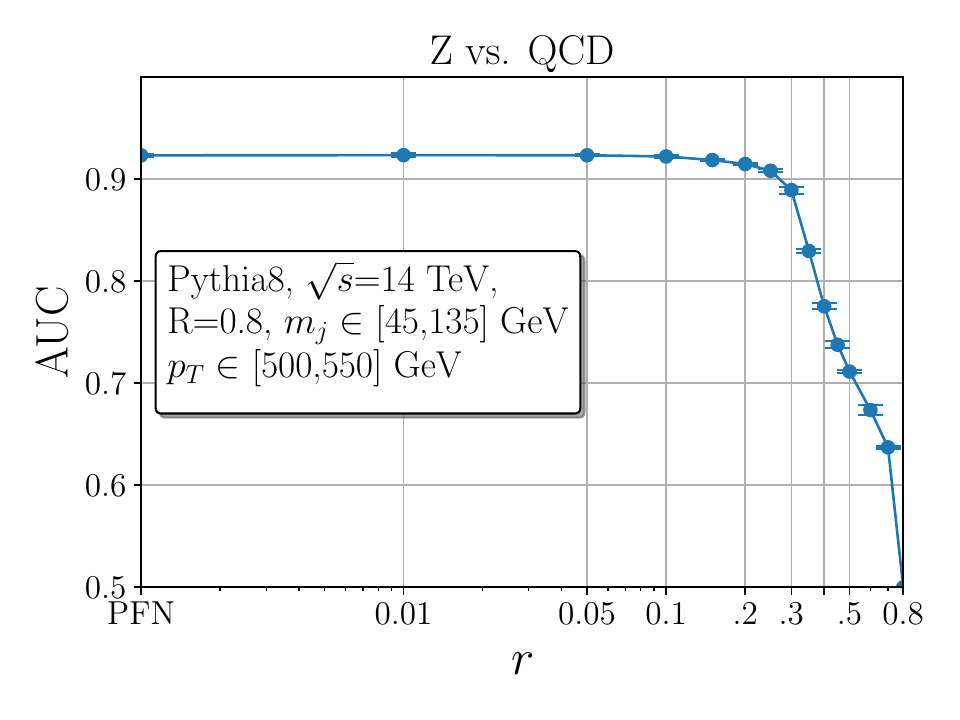} 
    \includegraphics[width=0.485\textwidth]{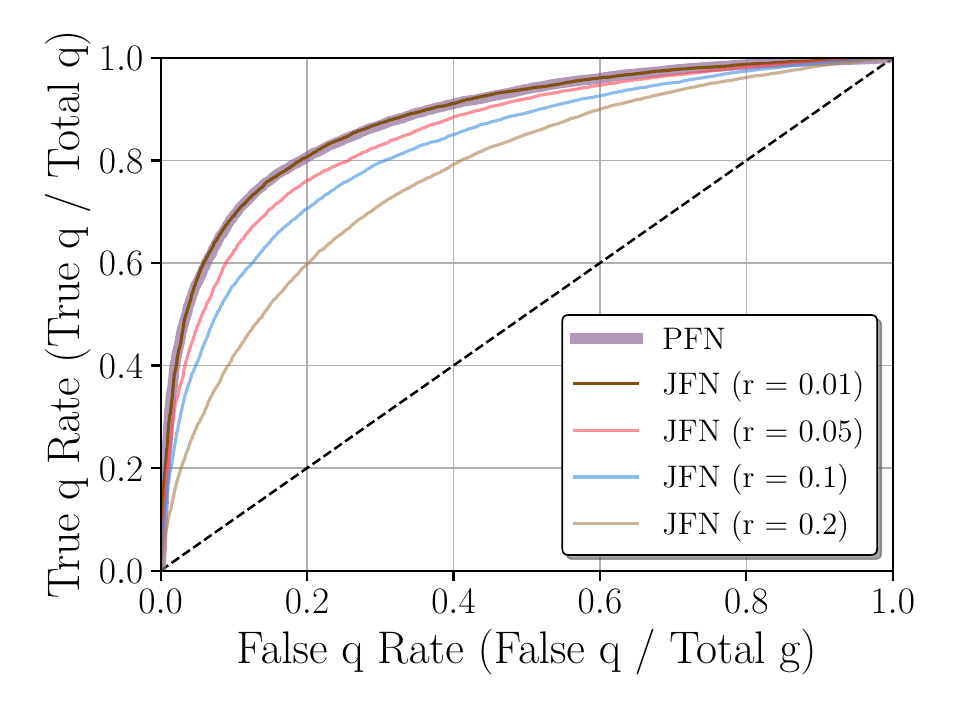}
    \includegraphics[width=0.485\textwidth]{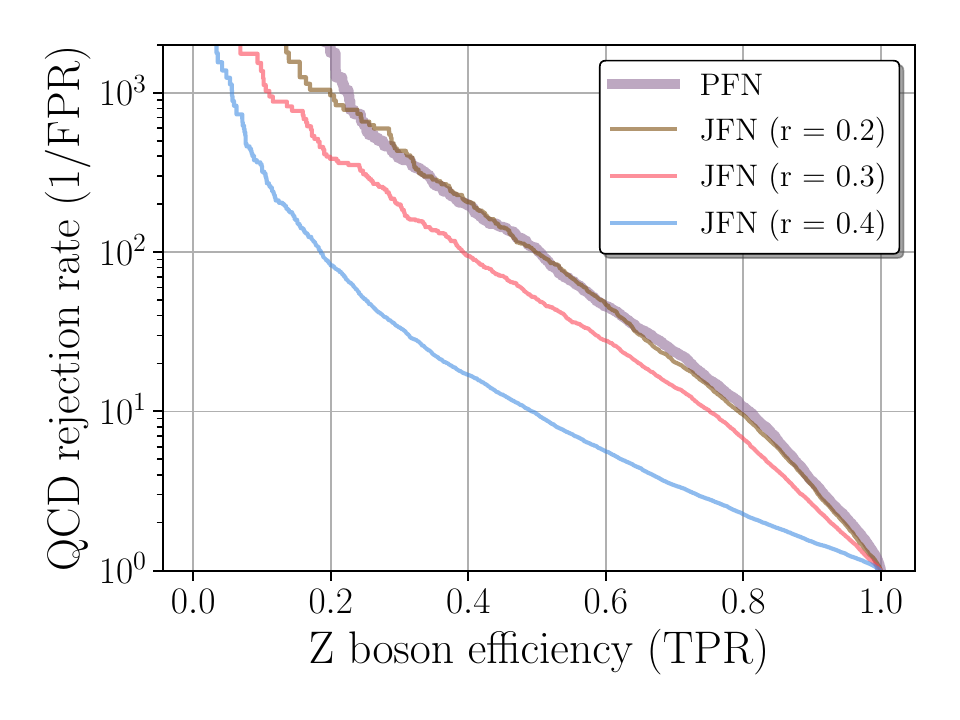}
    \caption{Top panel: AUC for quark vs. gluon (left) and Z vs. QCD (right) jet tagging using JFNs with different values of the (inclusive) subjet radius $r$. The PFN classifier is shown for reference at the leftmost value of $r$. Bottom panel: ROC curves for quark vs. gluon (left) and $Z$ vs. QCD jet tagging using JFNs and the PFN with different values of the (inclusive) subjet radius $r$ for the same datasets as the upper panel.~\label{fig:aucandroc_pt500}}
\end{figure}

Fig.~\ref{fig:aucandroc_pt500} (left) shows the JFN results for $q$ vs. $g$ jet discimination using inclusive subjet clustering and 2M jets for training, validation and testing. The top panel shows the AUC performance as we change the inclusive subjet radius $r$, and the bottom panel shows the ROC curves for several $r$. For comparison, we also show the PFN result. In the case of the AUC plot, we display the PFN classifier as the leftmost point on the $r$ axis. As expected, for $r=R$ we obtain a random classifier since all particles are clustered into a single subjet that is identical to the original jet. Similarly, we find that for the smallest subjet radii $r$ the performance of the PFN is recovered. Strikingly, however, the performance of the JFN does not diminish as $r$ is increased for values of the subjet radius $r \lesssim 0.01$. At this critical $r$ value, we have on average $n_{\rm subjets}/n_{\rm hadrons}\approx 0.75$. This observation is corroborated by the ROC curve in the lower panel, which shows that there is no performance loss in the JFN $(r=0.01)$ as compared to the PFN.
This demonstrates that there is little-to-no information encoded in the very collinear emissions relevant for discriminating $q$ vs. $g$ jets, and suggests that collinear safe inputs are sufficient for the purpose of $q$ vs. $g$ classification. In section \ref{sec:learning}, we will further discuss the physical interpretation of this critical $r$ value.

\begin{center}
\begin{table}[t]
\centering
\begin{tabular}{|c|c|cl|c|c|}
\cline{1-2} \cline{5-6}
\textbf{Model} & \textbf{AUC q vs. g} & \multicolumn{2}{c|}{\multirow{9}{*}{}} & \textbf{Model} & \textbf{AUC $Z$ vs. QCD} \\ \cline{1-2} \cline{5-6} 
PFN                     & 0.8912 $\pm$ 0.0005           & \multicolumn{2}{c|}{}                  & PFN            & 0.9235 $\pm$ 0.0015    \\ \cline{1-2} \cline{5-6} 
JFN ($r=$ 0.005)            & 0.8911 $\pm$ 0.0002           & \multicolumn{2}{c|}{}                  & JFN ($r=$ 0.01)    & 0.9237 $\pm$ 0.0018    \\ \cline{1-2} \cline{5-6} 
JFN ($r=$ 0.01)             & 0.8904 $\pm$ 0.0009           & \multicolumn{2}{c|}{}                  & JFN ($r=$ 0.05)    & 0.9236 $\pm$ 0.0010    \\ \cline{1-2} \cline{5-6} 
JFN ($r=$ 0.015)            & 0.8865 $\pm$ 0.0011           & \multicolumn{2}{c|}{}                  & JFN ($r=$ 0.1)     & 0.9227 $\pm$ 0.0017    \\ \cline{1-2} \cline{5-6} 
JFN ($r=$ 0.02)             & 0.8812 $\pm$ 0.0008           & \multicolumn{2}{c|}{}                  & JFN ($r=$ 0.15)    & 0.9189 $\pm$ 0.0009    \\ \cline{1-2} \cline{5-6} 
JFN ($r=$ 0.05)             & 0.8550 $\pm$ 0.0004           & \multicolumn{2}{c|}{}                  & JFN ($r=$ 0.2)     & 0.9150 $\pm$ 0.0012    \\ \cline{1-2} \cline{5-6} 
JFN ($r=$ 0.1)              & 0.8207 $\pm$ 0.0009           & \multicolumn{2}{c|}{}                  & JFN ($r=$ 0.3)     & 0.7755 $\pm$ 0.0025    \\ \cline{1-2} \cline{5-6} 
JFN ($r=$ 0.2)              & 0.7629 $\pm$ 0.0013           & \multicolumn{2}{c|}{}                  & JFN ($r=$ 0.4)     & 0.7115 $\pm$ 0.0015    \\ \cline{1-2} \cline{5-6} 
\end{tabular}
\caption{Numerical values for the AUC for both classification tasks considered in this section.~\label{tab:AUC}}

\end{table}    
\end{center}

Fig.~\ref{fig:aucandroc_pt500} (right) shows the analogous JFN results for $Z$ vs. QCD jet classification using inclusive subjet clustering and 500k jets for training, validation and testing. 
We observe again that the JFNs smoothly converge to the result of the PFN. Different than for quark vs. gluon jet tagging, we can now choose a significantly larger subjet radius $r\lesssim 0.1$ without compromising the performance of the classifier. 
This is related to the fact that in this case, the boosted $Z$-boson decay products generally lead to a two-pronged jet substructure, whereas QCD (quark and gluon) jets exhibit a single-pronged jet substructure (see Fig.~\ref{fig:tau2tau1} as well as Refs.~\cite{Larkoski:2017jix,Marzani:2019hun} for different observables and perturbative calculations that characterize the radiation patterns of QCD and boosted $Z$ jets). In general, machine-learned classifiers can make use of more information than the one- vs. two-pronged structure inside these jets; for $r\sim 0.1$, a significant fraction of the hadrons inside the jets are clustered into subjets, $n_{\rm subjets}/n_{\rm hadrons}\approx 0.4$. However, due to the observed saturation up to $r\sim 0.1$, we conclude that the information contained in collinear emissions is significantly less relevant for this classification task compared to quark vs. gluon jet tagging. This is due to the physical scales that are relevant for the different jet classification tasks, which we will explore in more detail in section~\ref{sec:learning}. For completeness, we list the numerical values for the AUC including uncertainties in table~\ref{tab:AUC}.

\subsection{Soft safety}

\begin{figure}[t!]
    \centering
    \includegraphics[width=0.495\textwidth]{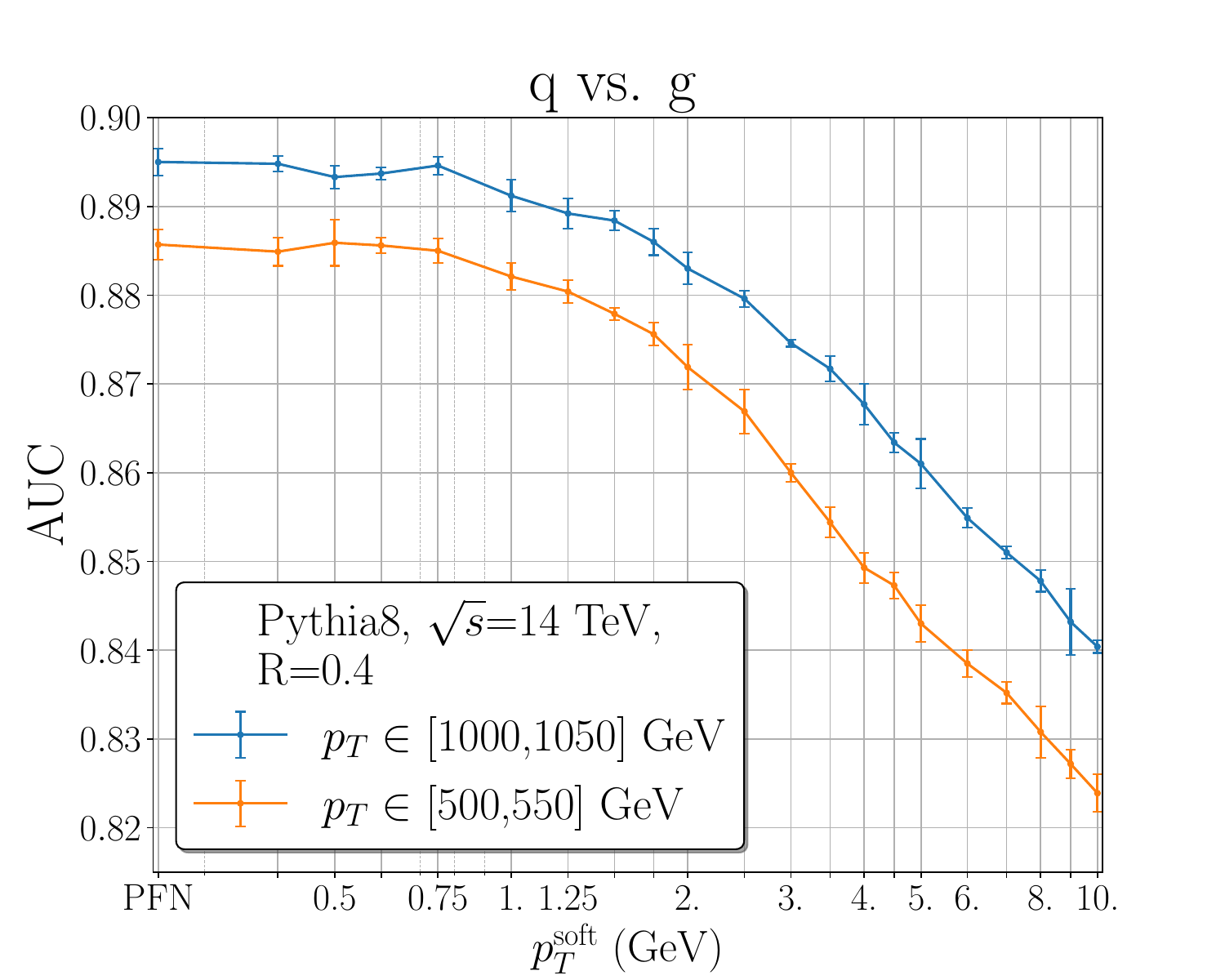} \includegraphics[width=0.495\textwidth]{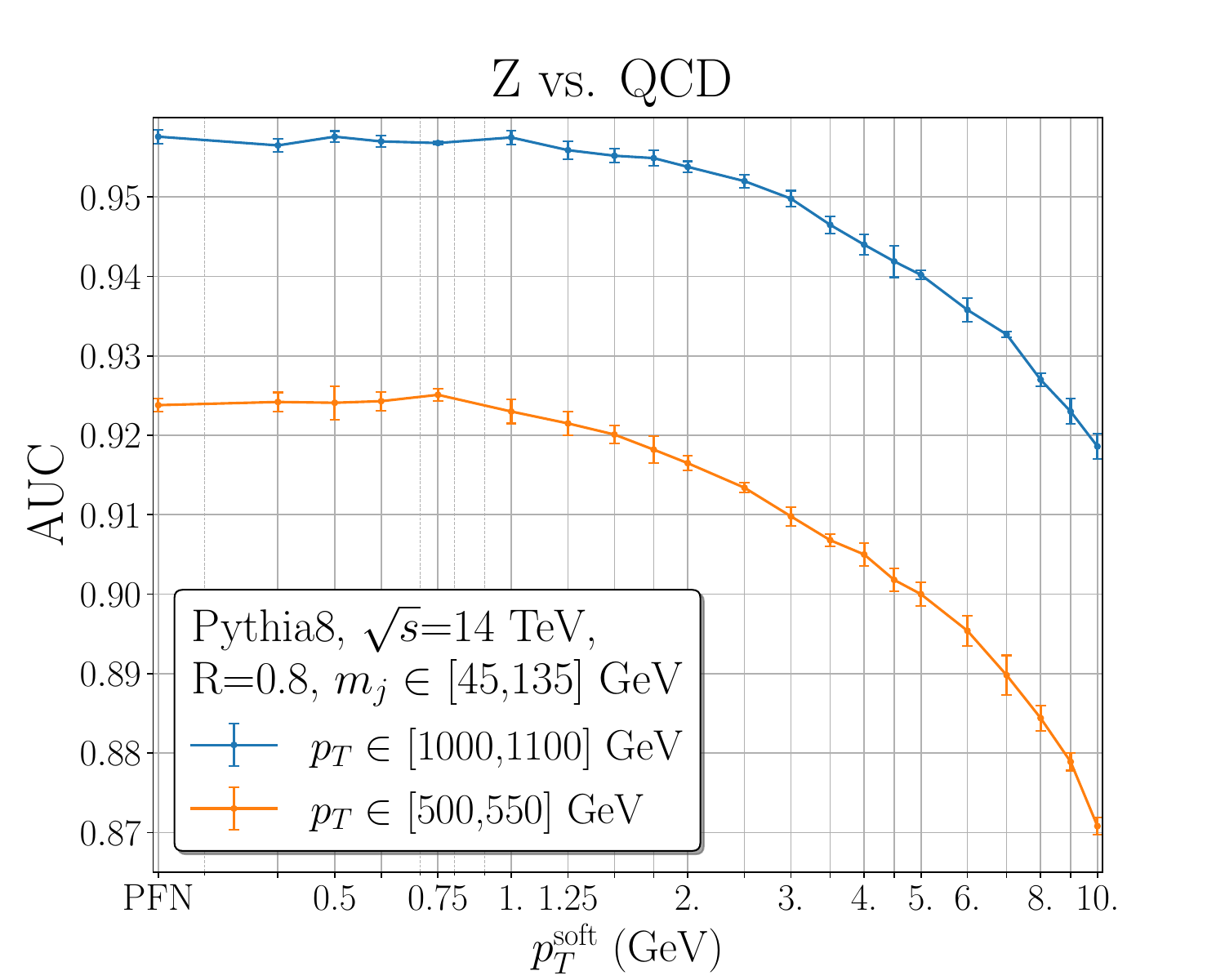} 
    \caption{AUC of the JFN as a function of the cut on the subjet's transverse momenta $p_T^{\rm soft}$ for quark vs. gluon (left) and $Z$ vs. QCD (right) jet tagging.~\label{fig:ptsoft}}
\end{figure}

Next, we are going to study the sensitivity of the classifier to soft emissions by analyzing the JFN performance as a function of the cut on the subjet's transverse momenta $p_T^{\rm soft}$. The results for the AUC are shown in Fig.~\ref{fig:ptsoft} for both quark vs. gluon and $Z$ vs. QCD jets. We studied the effects of the soft cut for two ranges of transverse momenta for the jet and using 500k jets as the dataset for each classification tasks. Again, we observe that the PFN performance is achieved for finite values of $p_T^{\rm soft}$ that are ${\cal O}(1~\text{GeV})$. We observe only small differences for the two classification tasks. The critical value for $p_T^{\rm soft}$ where the performance reaches a plateau is independent of the total jet transverse momentum $p_T$ within the displayed errors. This indicates that there is a soft scale below which no further information is added that can improve the classification performance. We conclude that IRC-safe information is indeed sufficient to carry out the two jet classification tasks considered in this work, which is consistent with theoretical considerations in the literature~\cite{Neill:2018uqw,Larkoski:2019nwj}. However, the relatively low values of $p_T^{\rm soft}$ that are needed to achieve the maximum performance indicate that nonperturbative effects are generally relevant. These observations may inform the design of suitable observables, which we leave for future work.

For both jet classification tasks, we have shown that for a range of subjet radii $r$ and $p_T^{\rm soft}$ cuts, the JFN exhibits no significant difference in performance compared to the PFN. That is, the JFN classifier here is ``gapless'' in the sense that we smoothly approximate the PFN performance (for finite values of $r$ and $p_T^{\rm soft}$). The clustering of soft and collinear emissions into subjets does not affect the performance as long as $r$ and $p_T^{\rm soft}$ are sufficiently small. This is in contrast to previous studies based on observables such as $N$-subjettiness variables, which exhibit a small but persistent performance gap to PFNs~\cite{Komiske:2017aww,Datta:2017rhs,Komiske:2018cqr}. The JFN provides the first example of a classifier with IRC-safe inputs that achieves equivalent performance to the IRC-unsafe PFNs for several classification tasks. Our results are consistent with the intuitive expectation that very low-energy particles are essentially uncorrelated with the hard process and therefore do not provide relevant information for typical classification tasks in high-energy physics. The main question of our paper has thus been answered by these observations. At least for the two classification tasks considered here, we have found that IRC-safe information is sufficient to close the gap to IRC-unsafe classifiers. This was achieved by using the machine learning architecture and input type (momentum, position information) for both cases and by including subjet reclustering as a preprocessing step in the IRC-safe case. We note that our conclusions come with the following caveat. While we are going to identify relevant physical scales with the performance of the classifiers, it is possible that future advances in machine learning lead to more powerful algorithms that may require us to reduce the subjet radius $r$ and cutoff $p_T^{\rm soft}$ to match the performance of IRC-unsafe classifiers. 

\section{Learning physical scales~\label{sec:learning}}

As discussed in the previous section, the performance of the JFNs based on IRC-safe input matches that of the IRC-unsafe PFNs for finite values of the subjet radius $r$ where a significant fraction of hadrons is clustered into subjets. In this section, we quantify in more detail the onset of the drop in performance when the subjet radius crosses certain physical scales. Here we only focus on inclusive instead of exclusive subjet reconstructions since we are primarily interested in the physical scale associated with a fixed value of subjet radius $r$. In order to identify the physical scale associated with the classification tasks and study its scaling behavior, we are going to analyze the AUC for different bins of jet transverse momentum $p_T$. Throughout this section, we only focus on the subjet radius $r$ and choose $p_T^{\rm soft}=0$~GeV since we already identified the relevant soft scale in the previous section.

\begin{figure}[t!]
    \centering
    \includegraphics[width=0.495\textwidth]{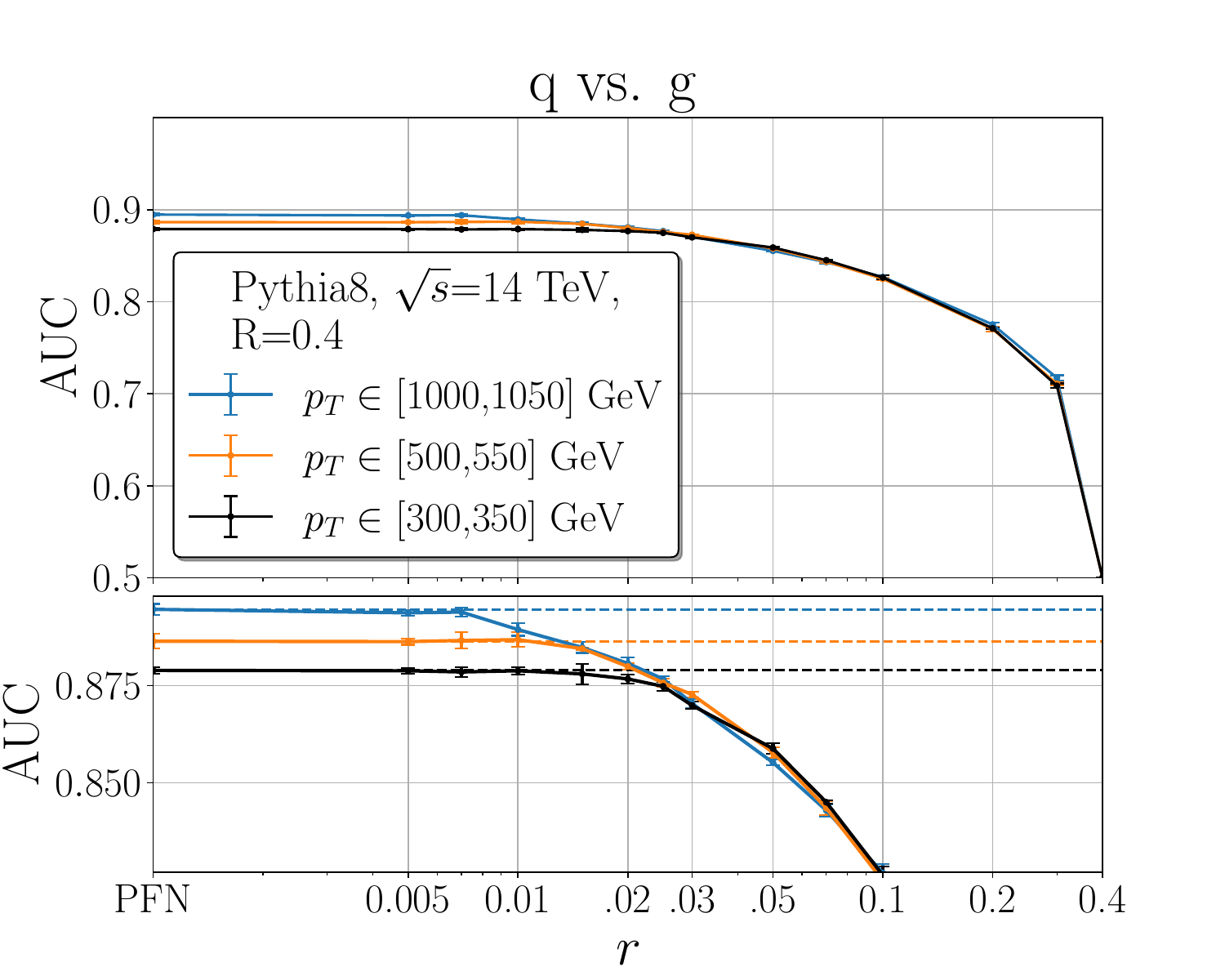}  
    \includegraphics[width=0.495\textwidth]{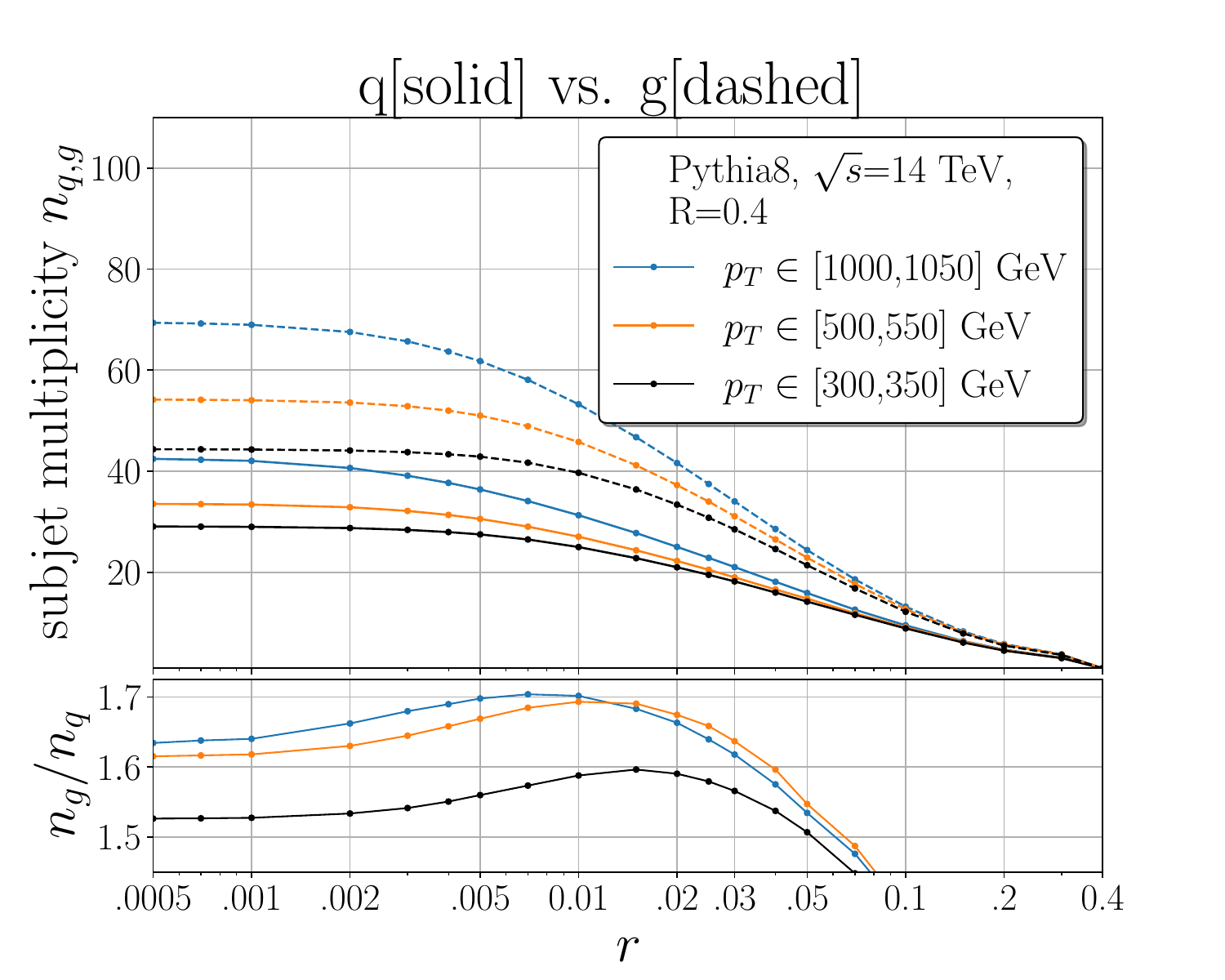}
    \caption{Left: AUC for quark vs. gluon jet classification for three different jet $p_T$ intervals as a function of the subjet radius $r$. Upper right: Average subjet multiplicity $n_{q,g}$ for quark (solid) and gluon (dashed) jets. Lower right: Ratio of the average subjet multiplicities $n_g/n_q$ for the three jet $p_T$ intervals.~\label{fig:auc_qvsg_pt300vs500vs1000}}
\end{figure}

In the upper left panel of Fig.~\ref{fig:auc_qvsg_pt300vs500vs1000}, we show the AUC for $q$ vs. $g$ discrimination for three different bins of the jet transverse momentum. For comparison, we show analogous to Fig.~\ref{fig:aucandroc_pt500} the PFN result as the left-most point ($r=0.001$). For all three classification tasks we used 500k jets for training, validation and testing. First, we notice that at sufficiently large $r$ all three AUC curves merge and the classification performance is (approximately) independent of the jet transverse momentum. This can be traced back to the approximate scale invariance of the QCD parton shower cascade. Second, we observe that in all three cases, the AUC reaches a plateau for finite values of the jet radius as $r$ is decreased.  In the lower left panel of Fig.~\ref{fig:auc_qvsg_pt300vs500vs1000}, we show the results in the transition region in more detail. The onset of the plateau shifts to the left as the jet $p_T$ is increased. For higher jet $p_T$, the jet constituents are more collimated leading to a smaller critical value $r$ where the JFNs match the PFN performance. We can identify the following approximate scale where the AUC reaches a plateau and agrees with the PFN results:
\begin{equation}
    p_T \cdot r\sim 5~\text{GeV}\,.
\end{equation}
Since there is no additional physical scale in the quark vs. gluon jet classification task besides the jet $p_T$, the agreement with the PFN result is achieved for relatively low energy scales. However, we would like to stress that the identified scale is still in the perturbative regime.  This suggests that because we only use particle momentum as our classifier's input that non-perturbative symmetries, such as isospin, forbid useful information at hadron level for discrimination. This would be broken and discrimination could improve if information sensitive to flavor was measured, which has been established in some recent studies of the jet charge \cite{Lee:2022kdn,Kang:2023ptt}.

Since the (IRC-unsafe) particle multiplicity is known to be a powerful discriminant for quark vs. gluon jet tagging~\cite{Larkoski:2019nwj, Frye:2017yrw, Gallicchio:2011xq}, we study the relation of our results to the average subjet multiplicities $n_{q,g}$ as a function of the subjet radius $r$, which is shown in the upper right panel of Fig.~\ref{fig:auc_qvsg_pt300vs500vs1000}. As expected, the subjet multiplicities for both quark and gluon jets increase as the subjet radius $r$ is decreased. In the limit $r\to 0$, the subjet multiplicity smoothly asymptotes to the particle multiplicity. The expected value for the ratio of the particle multiplicities at leading order is $n_g/n_q = C_A / C_F = 9/4$ ~\cite{Brodsky:1976mg, Konishi:1978yx}. In agreement with the discussion in the previous section, we notice that the subjet radius where the quark vs. gluon jet AUC reaches a plateau is larger than the $r$ value at which the subjet multiplicity reaches the particle multiplicity by an order of magnitude. This confirms that matching the PFN performance with JFNs is a non-trivial result. As shown in the lower right panel of Fig.~\ref{fig:auc_qvsg_pt300vs500vs1000}, we observe that the ratio $n_g/n_q$ peaks at intermediate values of $r$, which is in the region of the better modeled perturbative physics~\cite{Gallicchio:2011xq, OPAL:1997dkk, ALEPH:1998pel, DELPHI:1998pdl}. Interestingly, the location of the peaks is approximately the same as where the AUC for quark vs. gluon jet tagging reaches the plateau and agrees with the PFN result. This interesting correlation indicates that we can increase the subjet radius $r$ without affecting the classification performance until the subjet multiplicity $n_g/n_q$ starts to decrease. 

\begin{figure}[t!]
    \centering
     \includegraphics[width=0.7\textwidth]{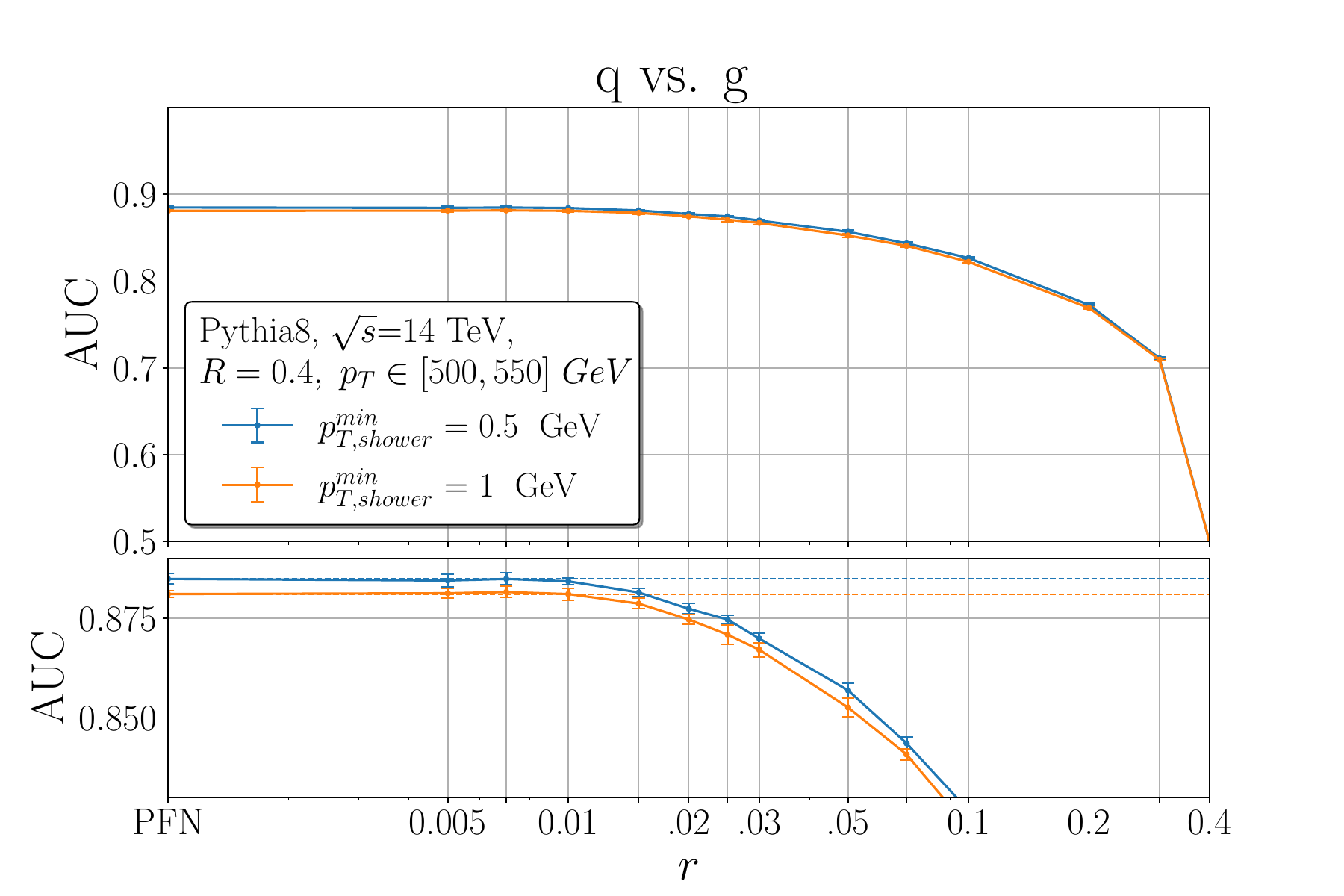}
    \caption{AUC for quark vs. gluon jet classification for two different choices of the perturbative shower cutoff parameter $p_{T,{\rm shower}}^{\rm min}$  in \textsc{Pythia} as a function of $r$. The value of $r$ where the maximum classification performance is achieved is independent of the shower cutoff within the displayed errors.~\label{fig:shower_cutoff}}
\end{figure}

Next, we investigate the dependence of our results on model parameters. In particular, we focus on the cutoff scale of the shower in \textsc{Pythia} at which the perturbative evolution ends and partons get converted to hadrons using the Lund string fragmentation model. The default value of this parameter is $p_{T{\rm shower}}^{\rm min}=0.5$~GeV. To assess the dependence on this parameter, we compare the default choice to $p_{T{\rm shower}}^{\rm min}=1$~GeV, which changes the transition from perturbative to non-perturbative splittings. The AUC results for the resulting classifiers are shown in Fig.~\ref{fig:shower_cutoff} for quark vs. gluon jets. We find that the asymptotic value of the AUC is slightly different when a higher transition scale is used. This is expected since the hadronization module used in \textsc{Pythia} masks differences between quark and gluon jets. More importantly, we observe that indeed the plateau is reached for the same finite value of $r \sim 0.01$, i.e. the scale we identified is independent of the choice of this parameter in \textsc{Pythia}. This illustrates that while the detailed modeling of soft physics is needed to describe the asymptotic AUC of the classifier, we observe that its peak performance is determined by a finite value of the subjet radius.

\begin{figure}[t!]
    \centering
     \includegraphics[width=0.7\textwidth]{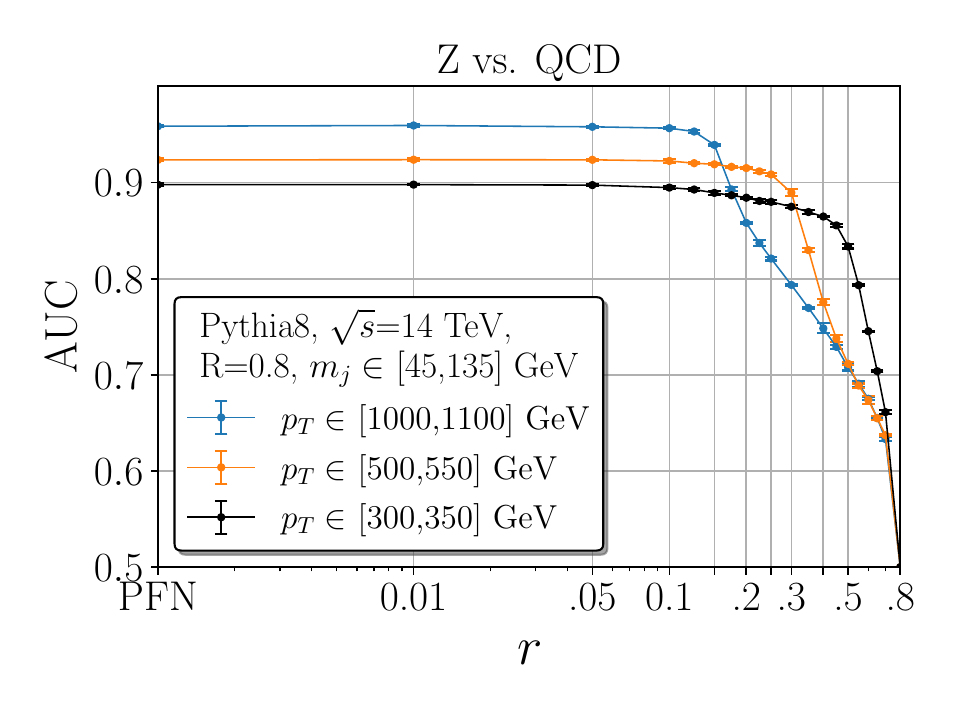}
    \caption{AUC for $Z$ vs. QCD jets for three different jet $p_T$ intervals as a function of the subjet radius $r$.~\label{fig:auc_ZvsQCD_pt300vs500vs1000}}
\end{figure}

Next, we consider QCD vs $Z$ jet classification. The AUC for three jet transverse momentum intervals is shown in Fig.~\ref{fig:auc_ZvsQCD_pt300vs500vs1000} as a function of the subjet radius $r$.  In all three cases we used 500k jets for training, validation and testing. As the jet $p_T$ is increased, the value of the subjet radius $r\sim 0.1-0.2$, where JFNs match the PFN performance is shifted to the left. This observation is generally consistent with quark vs. gluon jet tagging discussed above. We note that for the AUC curve with $p_T\in[300,350]$~GeV the choice of the jet radius $R$ might start to play a role. The value of $r$ where the performance reaches the plateau is roughly a factor of 10 higher compared to quark vs. gluon jet tagging. As already hinted at above, this is due to the presence of different physical scales. QCD jets do not have any additional intrinsic scales except for the hadronization scale. Instead, jets that contain the decay products of the boosted $Z$ boson are sensitive to the $Z$-boson mass $M_Z$.

\begin{figure}[t!]
\centering
\includegraphics[width=0.6\textwidth]{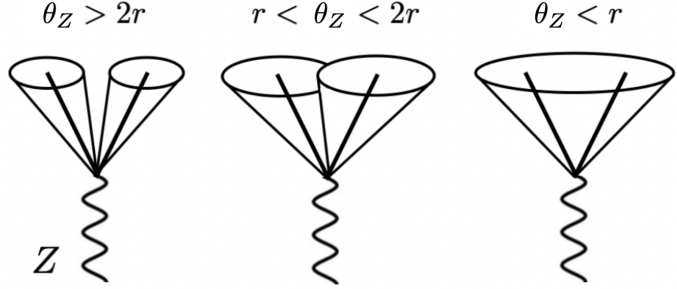}
\caption{Reclustering of the $Z$-boson decay products into subjets with different radii.~\label{fig:Zqq}}
\end{figure}

In order to gain further insights into the underlying physics, we are going to study the distribution of the opening angle of the two leading subjets $\theta_{12}$. This variable is closely related to the 2-pronged structure of $Z$ jets and serves as a useful discriminant since at leading order the $Z$-boson decays into a quark and anti-quark, which correspond to the two leading subjets at this order. The boosted $Z$ decay products have an opening angle $\theta_Z$, which is determined by $M_Z$ and the jet transverse momentum $p_T$ as
\begin{equation}
    \theta_{Z} \sim \frac{2M_Z}{p_T} \,.
\end{equation}
For higher $p_T$, the decay products are more boosted and $\theta_Z$ is smaller. See Fig.~\ref{fig:Zqq} for an illustration of the boosted $Z$-boson decay products clustered into subjets. If the subjet radius parameter is sufficiently small $r<\theta_{Z}/2$, the $Z$-boson decay products are clustered into separate subjets. Instead, for $r>\theta_{Z}$, they are merged into a single subjet. In the intermediate region, $r<\theta_{Z}<2r$, they are identified as two separate subjets but the subjet catchment areas overlap. 

\begin{figure*}[t!]
\includegraphics[width=0.325\textwidth]{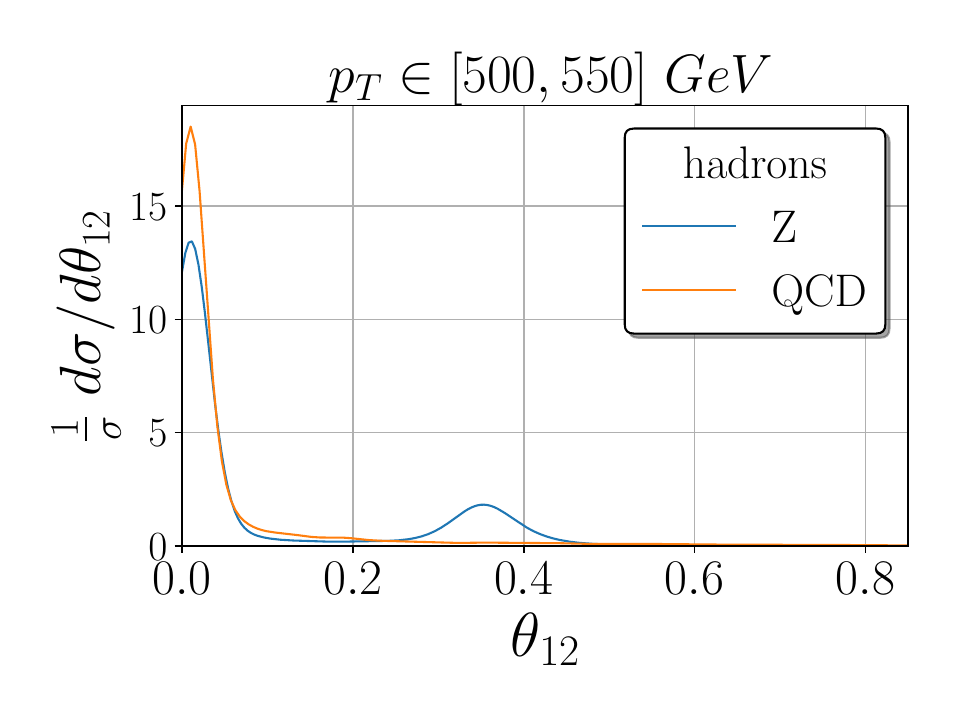}
\hspace*{.00cm}
\includegraphics[width=0.325\textwidth]{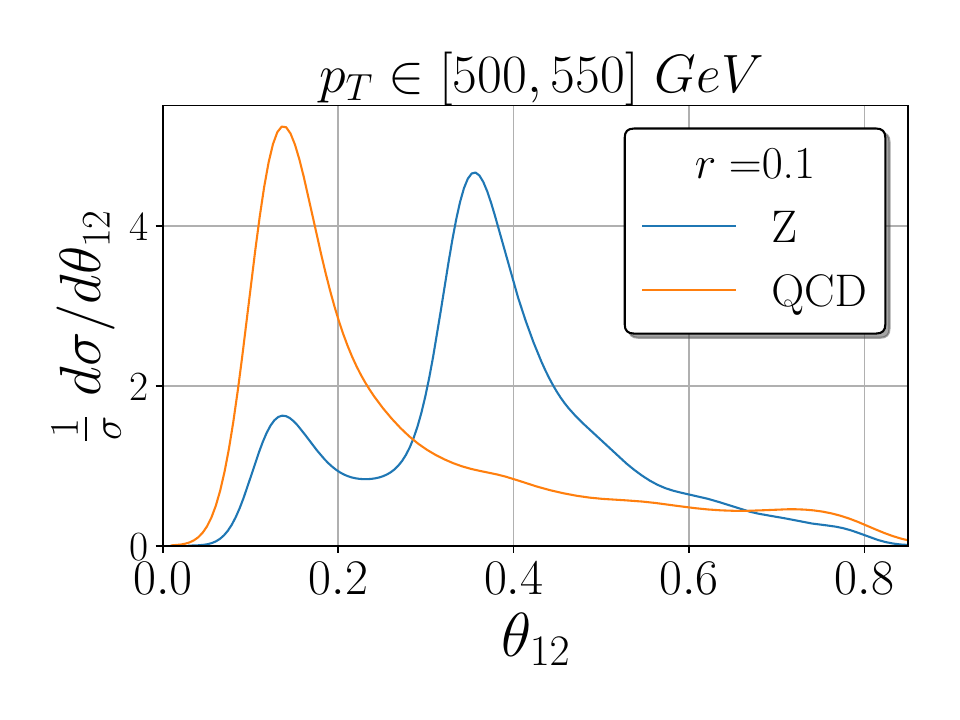}
\hspace*{.00cm}
\includegraphics[width=0.325\textwidth]{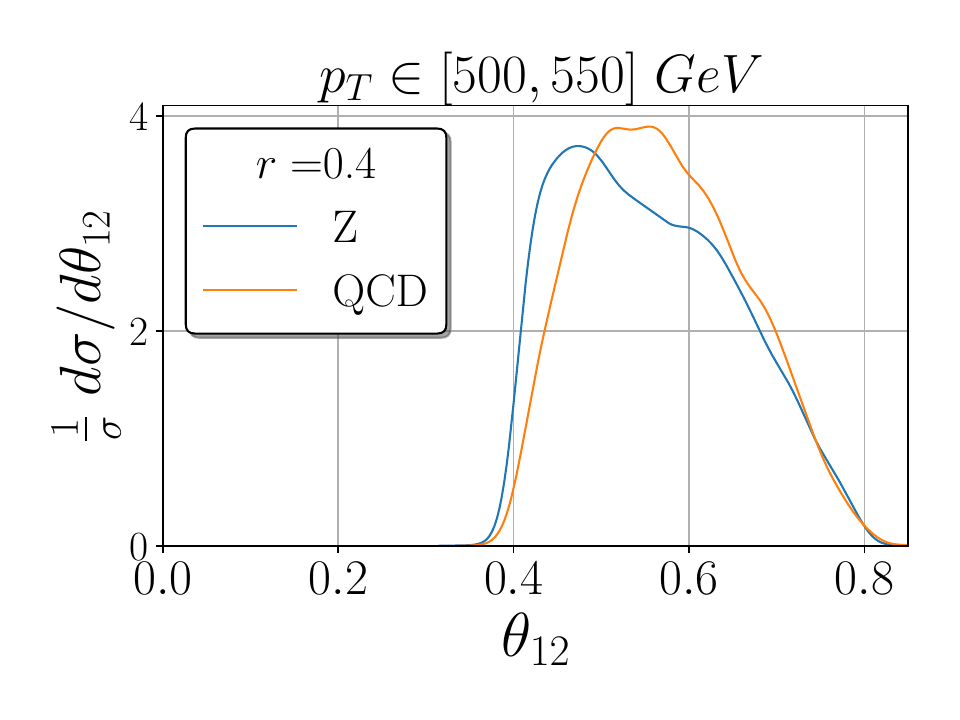}
\includegraphics[width=0.325\textwidth]{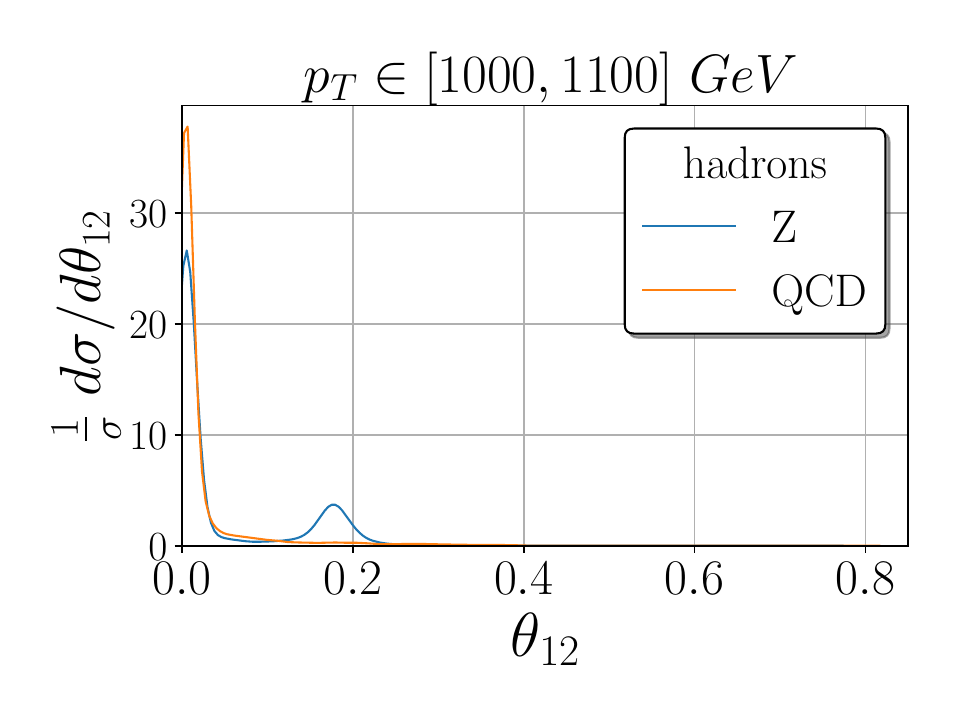}
\hspace*{.00cm}
\includegraphics[width=0.325\textwidth]{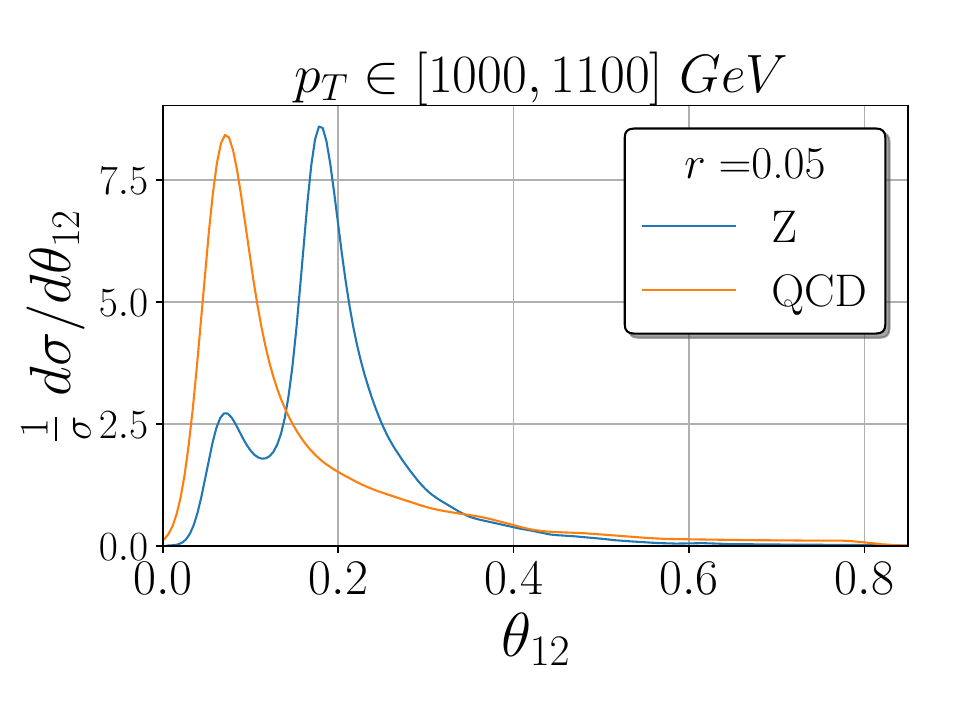}
\hspace*{.00cm}
\includegraphics[width=0.325\textwidth]{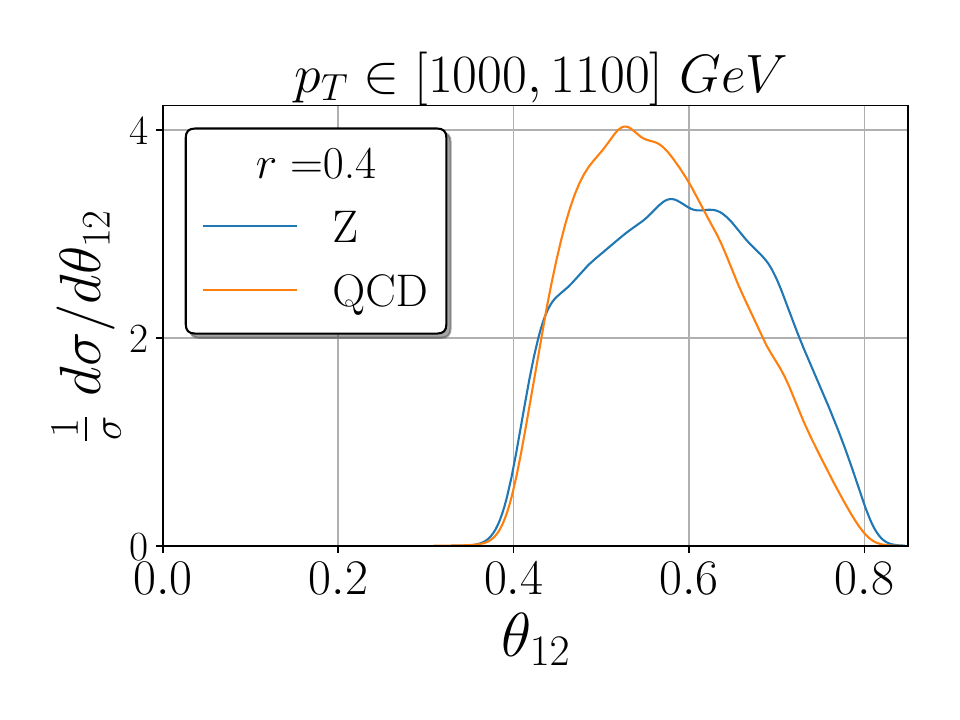}
\caption{Distributions of the opening angle $\theta_{12}$ between the two leading subjets for both QCD and $Z$ jets. We show the results for the two leading hadrons ($r=0$, left column) and two representative $r$ values (middle and right column). The upper and lower row correspond to two intervals of the jet transverse momentum $p_T$.~\label{fig:energies_angles}}
\end{figure*}

In Fig.~\ref{fig:energies_angles}, we show the distributions of the opening angle $\theta_{12}$ between the first two leading subjets for both QCD and $Z$ jets for different values of the subjet radius $r$. Here, $\theta_{12}$ corresponds to the geometric distance in the $\eta$-$\phi$ plane, i.e. without rescaling by the jet radius $R$.  The left column of Fig.~\ref{fig:energies_angles} shows the opening angles $\theta_{12}$ between the two leading hadrons, which corresponds to the subjet radius $r=0$. The middle column shows the $\theta_{12}$ distributions for $r$ values in the plateau region where the AUC of the JFNs matches the PFN result. The right column corresponds to higher $r$ values where the AUC has dropped significantly, see Fig.~\ref{fig:auc_ZvsQCD_pt300vs500vs1000} above. The top and bottom row correspond to two different jet $p_T$ intervals as indicated in the figure. We observe that for both QCD and $Z$ jets, the distributions are bounded from below by the chosen subjet radius, $\theta_{12}>r$, and the distributions vanish when the angle between the leading subjets reaches the jet radius $\theta_{12}\lesssim R=0.8$. Due to collinear QCD emissions, the distribution of the angle between the two leading hadrons peaks at $\theta_{12}\sim 0$. As the subjet radius is increased, it peaks close to the lower bound $\theta_{12}\sim r$. Eventually, the $\theta_{12}$ distribution becomes broader for large values of $r$. Instead, both at hadron level and for $r$ values in the plateau region of the AUC, the $\theta_{12}$ distribution of $Z$-jets has a two-peak structure. The left peak is due to QCD emissions and it occurs at the same $\theta_{12}$ value as the single peak of QCD jets. The second peak occurs around the opening angle of the $Z$ decay products $\theta_{12}\sim\theta_Z$. The width of the peaks scales as $\sim 1/p_T$. When $r$ is chosen in the plateau region of the AUC in Fig.~\ref{fig:auc_ZvsQCD_pt300vs500vs1000}, the JFN performance agrees with the PFN result. In this region, the two-peak structure of the $Z$-jet $\theta_{12}$ distribution can be clearly identified. The two-prong structure of $Z$ jets is the most prominent feature that distinguishes the two jet samples and it is clearly resolved as long as $r$ is sufficiently small. In this region, we found that the JFN performance is the same as the IRC-unsafe PFN result. 
While the location of the $\theta_{12}\sim\theta_Z$ peak is fixed, the peak due to QCD emissions moves to larger $\theta_{12}$ as $r$ is increased. Eventually, the two peaks start to merge. This is illustrated in the right-most column of Fig.~\ref{fig:energies_angles}, which shows the $\theta_{12}$ distributions for $r$ values where the JFN performance is significantly below its maximal value. The two peaks of the $Z$ jets have merged into one peak and the distribution is very similar to QCD jets. In this case, the $Z$-decay products cannot be clearly resolved and the performance of the classifier deteriorates. At this scale, the classifier does not have access to the UV physics anymore and as such the performance for the $p_T=[1000,1100]$~GeV jets matches the performance for the $p_T=[500,550]$~GeV jets. By comparing the upper and lower row of Fig.~\ref{fig:energies_angles}, we observe that the location of the peaks is shifted to lower values for higher jet $p_T$. In addition, the width of the peaks is narrower $\sim 1/p_T$. This agrees with the observation that for higher jet $p_T$, the end of the AUC plateau in Fig.~\ref{fig:auc_ZvsQCD_pt300vs500vs1000} is reached for smaller $r$ values.

\begin{figure}[t!]
\centering
\includegraphics[width=0.7\textwidth]{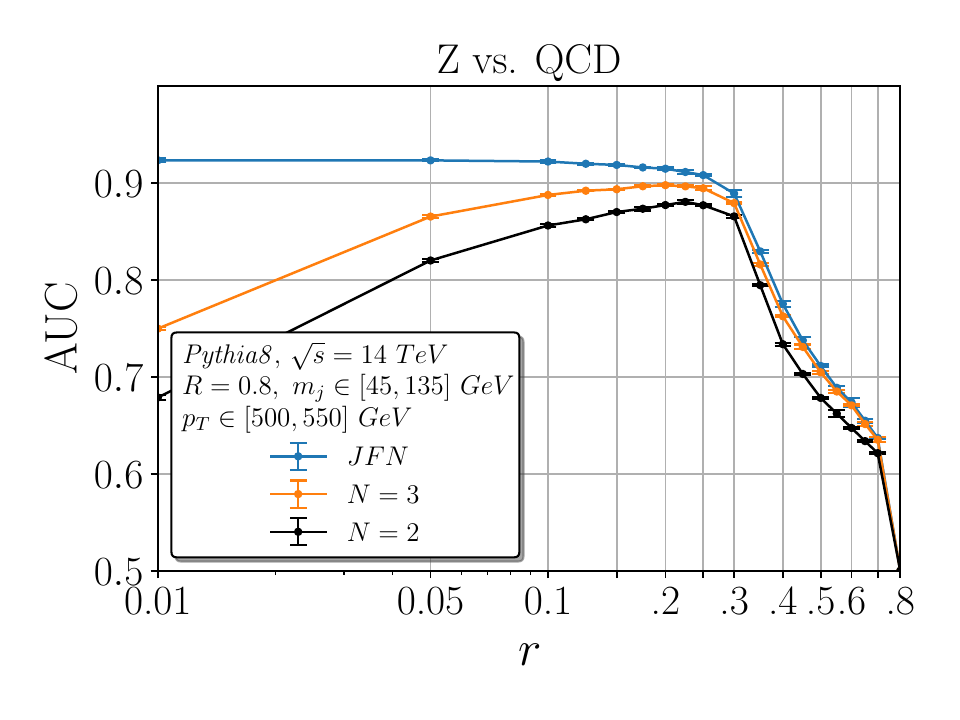}
\caption{AUC of the JFNs for $Z$ vs. QCD jets trained on the full information (inclusive subjets) compared to deep sets trained only on the two or three leading subjets.~\label{fig:N2vs3vsJFN}}
\end{figure}

Another way of illustrating the importance of resolving the $Z$-boson decay products is by training deep sets using only the information of the first few leading subjets. In Fig.~\ref{fig:N2vs3vsJFN}, we show the AUC for QCD vs $Z$ jets as a function of the subjet radius $r$ for three classifiers. We compare the JFNs to deep sets trained on the kinematic information of only the first two or three leading subjets. As an example, we use the jet transverse momentum interval of $p_T=[500,550]$~GeV. We observe that for large and intermediate values of the subjet radius $r$, the JFN performance is close to the deep sets trained on only two or three leading subjets. For small values of $r$, the leading two or three subjets do not contain enough information to match the JFN result. Especially, using the information of three leading subjets closely approximates the JFN performance down to a subjet radius of $r\sim 0.2$. The relevance of the third leading subjet corroborates the results of Ref.~\cite{Datta:2017lxt}, where the leading emission off the color dipole was identified as an important component for $Z$ vs. QCD jet classification.

Analogous classification tasks where physical scales can likely be identified are light QCD vs. $c$ or $b$-jets~\cite{ATLAS:2019bwq,ATLAS:2021cxe,ATL-PHYS-PUB-2020-014}, QCD vs. Higgs~\cite{Khosa:HiggsTagging} or QCD vs. top quark jets~\cite{Dreyer:2020brq}. We leave the exploration of these topics for future work.

\section{Performance vs. generalizability~\label{sec:generalization}}

Machine learning-based classifiers are often deployed in experimental analyses to tag jet topologies. A typical method is to train the classifier using fully supervised learning on precise theoretical simulations and apply it to experimental data~\cite{ATLAS:2022qby, ATLAS:2021otu, ATL-PHYS-PUB-2020-014}. However, this approach introduces model dependence as simulations do not perfectly match the actual data. In this section, we will explore some of the systematic uncertainties associated with this method. Other options that have been proposed include semi- or weakly-supervised techniques~\cite{Dery:2017fap,Metodiev:2017vrx}, as well as data-driven methods~\cite{Metodiev:2018ftz}.

When using fully supervised learning to develop classifiers, it is crucial to ensure that the model can generalize well to the unseen experimental data. For JFNs, soft and collinear particles are clustered into subjets making them less sensitive to the modeling of IR physics. Since it is generally challenging to model the very soft physics (both perturbative and nonperturbative) of collider events in universal Monte Carlo event generators, JFNs may have an advantage compared to PFNs in terms of generalizability. On the other hand, if too many particles are clustered into few subjets, the overall performance deteriorates. In order to assess whether a classifier performs well on unseen data, we train PFNs and JFNs with different parameters on \textsc{Pythia}~\cite{Sjostrand:2014zea} (training + validation data set) and test on \textsc{Herwig}~\cite{Bellm:2019zci} simulations. Here, \textsc{Herwig} can be considered as a surrogate for experimental data. It has been often observed that Pythia and Herwig envelope many jet substructure observables, sometimes referred to as the “Pythia-Herwig sandwich”. See e.g. Ref.~\cite{Kogler:2018hem}. We note that while the final results of both event generators are quite similar, the underlying physics of both the perturbative parton shower and the hadronization model can differ significantly. One generally expects that quark jets are quite similar in \textsc{Pythia} and \textsc{Herwig} but the results for gluon jets tend to differ more significantly~\cite{SchwartzCNN:2016rsd,Gras:2017jty,ATLAS:2014vax}. See also Ref.~\cite{Komiske:2016rsd}, where \textsc{Pythia} and \textsc{Herwig} studies were presented using Convolutional Neural Networks (CNNs). Moreover, in Ref~\cite{Butter:2022xyj} mixed \textsc{Herwig/Pythia} samples were used together with a Bayesian Network in order to increase model robustness. 

\begin{figure*}[t!]
\centering
\includegraphics[width=0.7\textwidth]{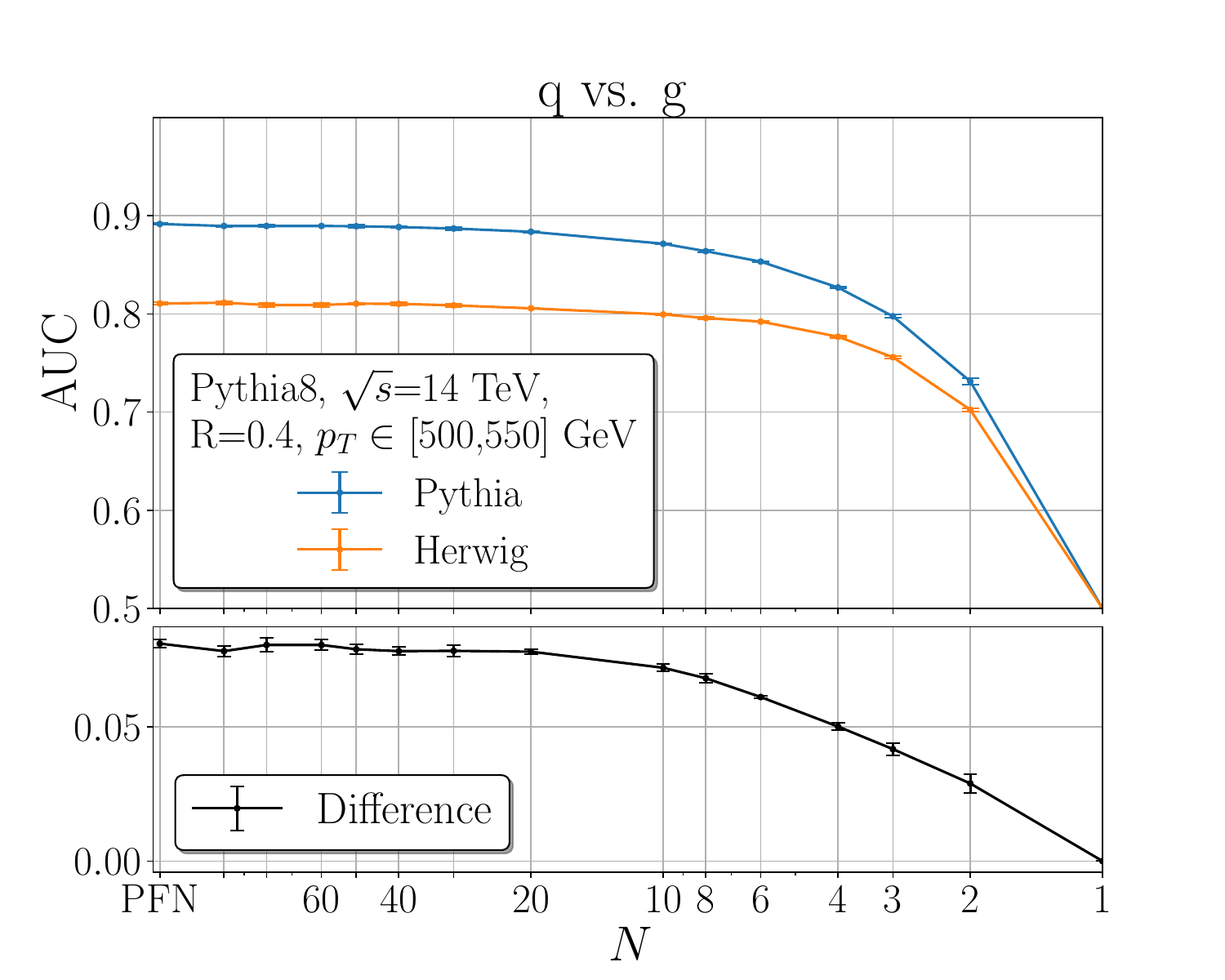}
\caption{Classification performance for quark vs. gluon jets using JFNs and exclusive $k_T$ clustered subjets plotted as a function of the number of subjets $N$. Upper panel: JFNs trained and tested on \textsc{Pythia}~\cite{Sjostrand:2014zea} (blue), JFNs trained on \textsc{Pythia} and tested on \textsc{Herwig}~\cite{Bellm:2019zci} (orange). Lower panel: The difference in the performance of the two results.~\label{fig:qg_auc_exc}}
\end{figure*}

We consider quark vs. gluon jet tagging for $p_T = [500,550]$~GeV using exclusive $k_T$ clustering of the subjets that are taken as input to the machine learning algorithm. Fig.~\ref{fig:qg_auc_exc} shows the AUC as a function of the number of the subjets $N$. Analogous to the previous figures we show the PFN result as the left-most point ($N=140$). The upper panel shows the result for JFNs as a function of $N$ trained on \textsc{Pythia} and tested on \textsc{Pythia} (blue) or \textsc{Herwig} (orange). In both cases, we observe a plateau in classifier performance as the number of (exclusive) subjets is increased. Within the shown errors, we observe that the AUC in both cases reaches its maximum value for $N\sim 30$. As expected, there is a performance gap when testing the \textsc{Pythia}-trained classifier on quark vs. gluon jets generated with \textsc{Herwig} compared to testing it on \textsc{Pythia} simulations. This observation is consistent with the results of Refs.~\cite{Komiske:2016rsd, Butter:2022xyj}. However, we observe that the performance gap decreases as $N$ decreases. To better visualize this aspect, we show in the lower panel the difference between the two AUC curves shown in the upper panel. The difference becomes smaller as $N$ is increased indicating improved generalizability of the model. Our findings suggest that clustering particles into subjets can reduce the overall performance, but it also masks modeling uncertainties of the IR physics leading to more robust classifiers. Interestingly, we find that the difference between \textsc{Pythia} and \textsc{Herwig} does not decrease for small $N$ for $Z$ vs. QCD jet classification (not shown). We do not observe an increased robustness using inclusive subjet clustering.
 
\begin{figure}[t!]
\centering
\includegraphics[width=0.7\textwidth]{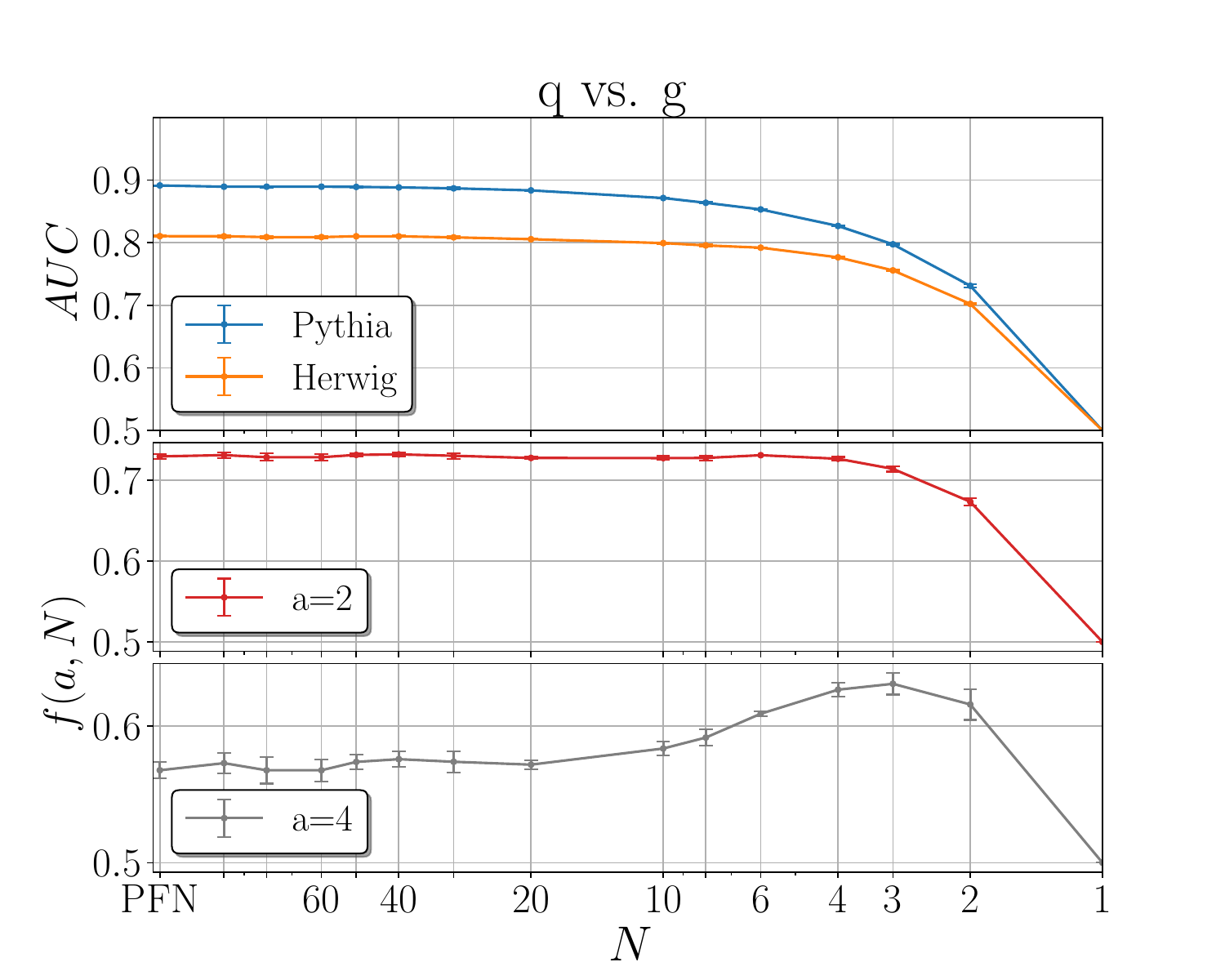}
\caption{Upper panel: AUC of the JFNs trained on \textsc{Pythia} and tested on either \textsc{Pythia} or \textsc{Herwig} plotted as a function of the number of (exclusive) subjets $N$, see also Fig.~\ref{fig:qg_auc_exc}. Middle and lower panels: The objective function $f(a,N)$ defined in Eq.~(\ref{eq:AUC_opt}), where $a$ is a weighting factor between optimal performance and generalizability.~\label{fig:optimization}}
\end{figure} 

We expect that the generalizability or robustness of machine learning-based classifiers will be useful for certain experimental applications where the trade-off between performance and generalizability needs to be considered. To illustrate this aspect, we introduce the objective function $f(a,N)$, defined as:
\begin{equation}\label{eq:AUC_opt}
    f(a,N) = {\rm AUC}_{\rm Pythia}(N) - a \cdot \left({\rm AUC}_{\rm Pythia}(N)- {\rm AUC}_{\rm Herwig}(N)\right)\,,
\end{equation}
where $N$ is the number of exclusive subjects. Here the performance and generalizability are combined additively and a weighting factor $a>0$ is introduced that allows us to increase/decrease the relevance of the two metrics. An optimization problem to find the optimal balance between performance (first term in Eq.~(\ref{eq:AUC_opt})) and generalizability (second term $\sim a$ in Eq.~(\ref{eq:AUC_opt})) can now be formulated as follows: For a given choice of the weighting factor $a$, find the maximal value of the objective function $f(a,N)$. The optimal number of exclusive subjets is then given by $N_{\rm opt} = \argmax_{N}f(a,N)$. We plot $f(a,N)$ for two different values of $a$ in Fig.~\ref{fig:optimization} (middle and lower panels). We observe that as $a$ is increased (the generalizability is weighted higher), the objective function peaks at an intermediate value of $N$. For example, for $a=4$ we find $N_{\rm opt}= 3$. While our objective function is constructed for illustration purposes, this result indicates that for certain experimental analyses that employ machine learning-based classifiers, it can be advantageous to use JFNs with a finite number of subjets to achieve the desired goals.

\section{Conclusions~\label{sec:conclusions}}

The classification of jets at collider experiments is relevant for a wide range of tasks in high-energy particle and nuclear physics. Over the past years, machine learning-based classifiers have been developed that can achieve impressive tagging performance. While machine learning generally outperforms traditional methods by efficiently making use of the full information content, it is often unclear where the performance difference is coming from. In particular, it had been unclear if classifiers based on infrared-collinear (IRC) information can match the performance of IRC-unsafe classifiers. IRC safety is primarily motivated by theoretical considerations ensuring that observables are tractable in perturbative QCD. In addition, it is expected that the very soft physics is uncorrelated to the hard partonic process making it unlikely to be the reason of the performance gap that has been observed between IRC-unsafe machine learning results and traditional IRC-safe observables. 

In order to address these questions, we introduced in this work a new family of classifiers, the Jet Flow Networks (JFNs). Here, particles inside a jet are first clustered into subjets and their position and momentum are taken as input to a permutation-invariant neural network (deep set). The clustering of subjets with a certain radius and transverse momentum cut allows us to control the sensitivity to soft and collinear emissions making the input to the classifier IRC safe. As the subjet radius (collinear safety) and momentum cut (soft safety) vanish, we recover the IRC-unsafe Particle Flow Networks (PFNs). We investigated both inclusive and exclusive subjet clustering, which can lead to important differences depending on the application. As representative examples, we considered two classification tasks: quark vs. gluon and $Z$ vs. QCD jet tagging. Interestingly, we observed that the JFN performance matches the IRC-unsafe PFN result for finite values of the subjet radius and the soft transverse momentum cut. This makes JFNs the first classifier based on IRC-safe input without a performance gap to their IRC-unsafe counterpart for several jet classification tasks. This observation answered the main question we aimed to address in this work and indeed IRC-safe information is sufficient for the jet classification tasks considered here. As the subjet radius is increased, the performance of the JFNs remains unchanged (and in agreement with the PFNs) until physical thresholds are crossed. For example, for quark vs. gluon jets this threshold is around 5~GeV, whereas for $Z$ vs. QCD jets it is determined by the kinematics of the hadronic boosted decay products of the $Z$-boson. The analogous threshold for the soft momentum cut is ${\cal O}(1~\text{GeV})$. This indicates that while the jet classification tasks are IRC safe, nonperturbative physics is generally relevant for jet classification. In addition, we found that JFNs may offer a decreased model dependence for certain classification tasks with only a modest tradeoff in performance. This was illustrated in section~\ref{sec:generalization} using a toy example and applications may depend on the type of experimental analysis that is carried out. This observation may lead to interesting applications of JFNs in collider phenomenology.

Our results shed new light on the information that machines learn in high-energy physics applications. As more powerful algorithms will be developed it will be interesting to revisit the question of the potential gap between classifiers based on IRC-safe and unsafe information. While more work is needed in this direction, our work represents an important step toward increasing the interpretability of machine learning methods in high-energy physics. In addition, we anticipate various applications of JFNs in heavy-ion collisions and the future Electron-Ion Collider~\cite{AbdulKhalek:2021gbh}.

\begin{acknowledgments}
We thank Giacinto Piacquadio for helpful discussions. DA is supported by the NSF Grant PHY2210533 and the Onassis Foundation. JM, MP are supported by the U.S. Department of Energy, Office of Science, 
Office of Nuclear Physics, under the contract DE-AC02-05CH11231. FR was supported by the Simons Foundation under the Simons Bridge program for Postdoctoral Fellowships at SCGP and YITP award number 815892; the NSF, award number 1915093; the DOE Contract No.~DE-AC05-06OR23177, under which Jefferson Science Associates, LLC operates Jefferson Lab, and in part by the DOE, Office of Nuclear Physics, Early Career Program under contract No. DE-SC0024358. AL is supported in part by the UC Southern California Hub, with funding from the UC National Laboratories division of the University of California Office of the President. This research used resources of the National Energy Research Scientific Computing Center (NERSC), which is supported by the Office of Science of the U.S. Department of Energy under Contract No. DE-AC02-05CH11231.

\end{acknowledgments}

\bibliographystyle{JHEP}
\bibliography{main.bib}

\end{document}